\newcommand \Mpc {h^{-1}{\rm Mpc}}

\newcommand \farcm{\hbox{$.\!\!^{\prime}$}}
\newcommand \arcm{\hbox{$^{\prime}$}}

\newcommand \kms {{\rm km~s}^{-1}}

\newcommand \beqn {\begin{equation}}
\newcommand \eeqn {\end{equation}}
\newcommand \Planck {{\em Planck }}

\newcommand \nczhecsred {10,589 }

\newcommand \nclusterhecsred {23 }

\newcommand \nclusterhecsredextended {121 }

\documentclass[apj]{emulateapj}
\usepackage{epsfig}
\usepackage{natbib}
\newcommand{\noprint}[1]{}

\begin{document}

\title{HeCS-red: Dense Hectospec Surveys of redMaPPer-Selected Clusters} 
\shorttitle{HeCS-red: Dense Hectospec Surveys of redMaPPer-Selected Clusters}
\shortauthors{Rines et al.}

\author{Kenneth J. Rines\altaffilmark{1}, Margaret J. Geller\altaffilmark{2}, 
Antonaldo Diaferio\altaffilmark{3,4}, Ho Seong Hwang\altaffilmark{5}, and Jubee Sohn\altaffilmark{2}} 
\email{kenneth.rines@wwu.edu}

\altaffiltext{1}{Department of Physics \& Astronomy, Western Washington University, Bellingham, WA 98225; kenneth.rines@wwu.edu}
\altaffiltext{2}{Smithsonian Astrophysical Observatory, 60 Garden St, Cambridge, MA 02138}
\altaffiltext{3}{Universita' di Torino, Dipartimento di Fisica, Torino, Italy}
\altaffiltext{4}{Istituto Nazionale di Fisica Nucleare (INFN), Sezione di Torino, Torino, Italy}
\altaffiltext{5}{Quantum Universe Center, Korea Institute for Advanced 
Study, 85 Hoegiro, Dongdaemun-gu, Seoul 02455, Republic of Korea}

\begin{abstract}

We use dense redshift surveys to explore the properties of galaxy clusters 
selected from the redMaPPer catalog of overdensities of red galaxies.
Our new survey, HeCS-red (Hectospec Cluster Survey of red-sequence selected clusters), 
includes \nczhecsred new or remeasured redshifts from MMT/Hectospec observations
of 
redMaPPer clusters at redshifts $z$=0.08-0.25 with large estimated richnesses (richness 
estimate $\lambda >64$).  Our spectra confirm that each of these candidate clusters 
corresponds to a large overdensity in redshift space.   
The redMaPPer photometric redshifts appear to have a slight bias towards higher redshifts, 
with $\overline{z_{spec}-z_{RM}}=-0.0028\pm0.0005$. 
We measure the scaling relation between velocity dispersion $\sigma_p$ and 
redMaPPer richness estimates $\lambda$.
The observed relation shows intrinsic scatter of 24\% in velocity dispersion at 
fixed richness, and a range of a factor of two in measured $\sigma_p$ at fixed richness.  
Several outliers from the relation have multiple structures along the line of sight.
We extend our analysis to HeCS-red-ext, a sample that includes 
several clusters selected by X-ray flux or Sunyaev-Zeldovich signal. 
The heterogeneous sample of \nclusterhecsredextended clusters in HeCS-red-ext shows similar
intrinsic scatter in the $\sigma_p-\lambda$ relation as the HeCS-red sample, but the range of 
$\sigma_p$ at fixed richness increases to a factor of three.  
We evaluate the membership probability estimates $P_{mem}$ for individual galaxies provided 
by redMaPPer.  
The spectroscopic membership fraction is larger than $P_{mem}$ for $0.05\leq P_{mem}\leq 0.7$; 
conversely, the spectroscopic membership fraction is smaller than $P_{mem}$ at $P_{mem}\geq 0.8$.
We compare richness 
estimates based on our spectroscopic samples to redMaPPer richness estimates 
and find good agreement on average, but a range of a factor of two in spectroscopic 
richness at fixed redMaPPer richness. 
Overall, within the high-richness and low-redshift cut of our sample, redMaPPer clusters correspond to 
genuine rich clusters.  Spectroscopically estimated parameters such as velocity 
dispersion correlate well with richness estimated from photometry, although 
the relations contain substantial scatter.

\end{abstract}


\section{Introduction}

Cosmological models 
make different predictions for the comoving number density
of clusters of fixed mass.  The evolution of cluster abundances 
depends strongly on the amount of dark matter and dark energy in the 
universe. Thus, many groups have used different cluster mass proxies to 
determine the mass function and constrain cosmological parameters \citep[e.g.,][and references
therein]{cirsmf,rines08,vikhlinin09b,henry09,mantz10a,rozo10,benson13,hasselfield13,planckszcosmo,planckszcatalog,planckszcosmo2015,mantz15,dehaan16}.  

The red sequence of cluster galaxies has been used by many 
investigators as a technique to discover new clusters \citep{gladders00,gladders05,koester07a}.
One of the most sophisticated algorithms to detect clusters with the 
red sequence is the redMaPPer algorithm \citep{rykoff14}.
\citet{rykoff14} apply the redMaPPer algorithm to imaging data
from the Eighth Data Release (DR8) of the Sloan Digital Sky Survey \citep[SDSS;][]{dr8}.
The redMaPPer cluster catalog includes an estimated richness 
that 
serves as a mass proxy.  Several tests show that 
redMaPPer richnesses correlate with other mass proxies, including
Sunyev-Zeldovich (SZ) signal \citep{rozo14d}, X-ray luminosity \citep{sadibekova14,rozo14d}, 
and weak lensing signal \citep{simet17a}.  However, these comparisons
usually are made either with samples of individual clusters selected
by their properties at other wavelengths (e.g., clusters with SZ detections 
in the \Planck catalog) or by stacking of clusters with the same 
richness parameter \citep[e.g.,][]{rykoff08,andreon14,simet17a}.   If the parameter
(e.g., richness) used to define cluster stacks has large intrinsic
scatter (e.g., in true cluster mass), then the ensemble cluster samples 
only provide signals averaged over a broad range of intrinsic cluster properties.

Here, we compare optical richness estimates to dynamical mass estimates 
based on dense redshift surveys of individual clusters.  Dynamical mass estimates 
have a long history beginning with \citet{zwicky1933,zwicky1937}.  
In numerical simulations, either the virial theorem or the caustic technique 
can provide cluster mass estimates with little bias but with some intrinsic 
scatter due to projection effects \citep{diaferio1999,evrard07,serra11,mamon13,gifford13,old14}.
Hydrodynamical simulations show that the velocity distribution of galaxies 
is very similar to that of dark matter particles \citep{faltenbacher06,lau10}, with the possible
exception of the brightest few galaxies \citep{lau10,wu13}.  Thus, virial masses, 
caustic masses, or dynamical mass proxies such as velocity dispersion 
are a powerful test of richness-based mass estimates.

We describe MMT/Hectospec spectroscopic observations of a sample of 27 clusters 
(6 observed previously) selected with redMaPPer richness parameter $\lambda>64$ and redshift $z=0.08-0.25$. 
The redshift surveys of these clusters test the impact of projection effects
on cluster identification and provide robust dynamical masses for comparison with the richness estimate 
$\lambda$. 
We extend our analysis to two additional redMaPPer clusters with Hectospec observations published here 
and to an additional 94 clusters with velocity dispersions from previously published wide-field optical spectroscopy.   

We discuss the cluster samples and spectroscopic data in $\S 2$.  We measure
the scaling relations in $\S 3$. We discuss
the implications of our results in the context of other cosmological observations
in $\S 4$.  
We assume a $\Lambda$CDM cosmology of
$\Omega_m$=0.3, $\Omega_\Lambda$=0.7, and $H_0$=100 $h$ km s$^{-1}$ Mpc$^{-1}$ for all calculations.

\section{Observations}

\subsection{Cluster Sample Selection}

Previous surveys \citep{cirsi,hecsultimate,hecsplanck} contain spectroscopy for several clusters in the
redMaPPer catalog.  However, these clusters were selected based on their 
X-ray or SZ properties.  As a result, scaling relations based on these samples 
could be biased relative to the scaling relations for a purely redMaPPer-selected 
sample.  

\begin{figure} 
\plotone{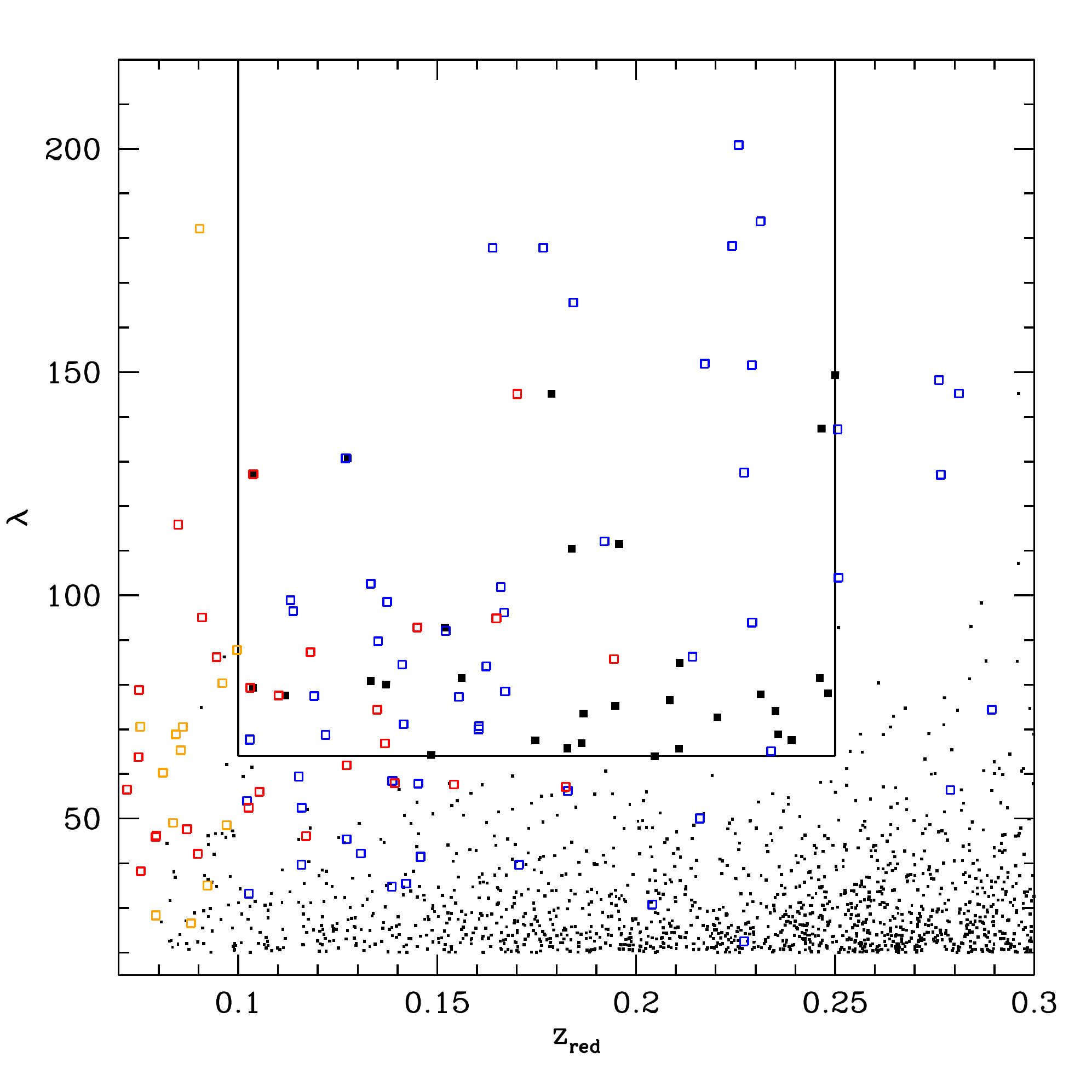}
\caption{\label{hecsredzlambda} redMaPPer richness estimates $\lambda$ versus redshift.  
Black squares show clusters within the spatial footprint of our redMaPPer-selected 
target sample.  Solid lines show the redshift and richness limits of the target sample.
Open blue, red, and orange squares show clusters from HeCS, HeCS-SZ, and CIRS respectively. 
}
\end{figure}

We define a redMaPPer-selected sample of 30 clusters 
covering redshift $0.10\leq z\leq 0.25$, 
redMaPPer richness parameter $\lambda \geq 64$, coordinates $\delta >10^\circ$,
and either $\alpha <9h$ or $\alpha >20h$ (Figure \ref{hecsredzlambda}).  
The redMaPPer catalog extends to $\lambda=20$ and $z=0.55$ \citep{rykoff14}; the sample we select
is limited to the richest clusters at low redshift.

Among the 30 clusters in HeCS-red, 1 cluster (A655) 
was observed with Hectospec for the Hectospec Cluster Survey \citep[HeCS;][]{hecsultimate} 
and 5 clusters (A586, A98S, A2409, A7, and A2443) were observed with 
Hectospec for  the Hectospec Cluster Survey of SZ-Selected clusters \citep[HeCS-SZ;][]{hecsplanck}. 
We observed 21 additional clusters for HeCS-red, yielding a redMaPPer-selected 
cluster sample of 27 clusters that is 90\% complete. 
Hereafter we call this sample of 27 redMaPPer-selected clusters HeCS-red.

We also observed one cluster, RMJ023054.9+024719.6
(with $\lambda=90.7\pm4.1$), 
from the declination range $0^\circ <\delta <10^\circ$.
We also include Hectospec data for A2355 ($\lambda=109.4\pm4.8$), 
originally selected based 
on its SZ signal from the {\em Planck} satellite \citep{planckszcatalog}.
Because even the central redshift of A2355 was poorly known at the time of 
observation \citep[e.g.,][lists a redshift of $z=0.1244$]{reflex}, spectroscopic targets for A2355
were selected solely by apparent magnitude with no color cuts. 
Hereafter, we call the extended sample including these two clusters and several 
from previously published work (see $\S 2.2.1$) HeCS-red-ext.

\subsection{Optical Photometry and Spectroscopy}

We use photometry primarily from the Tenth Data Release 
(DR10) of SDSS \citep{sdssdr10} 
to identify targets for spectroscopic observations.    
For a few clusters, inspection of DR10 photometry revealed that the catalogs 
contained 'holes,' typically around a bright galaxy or star where the background 
subtraction could be problematic.  For several of these holes, we used DR7 
photometry (which was less aggressive about removing regions around bright 
stars and galaxies) to select targets for spectroscopy. 
We acquired spectra with the
Hectospec instrument \citep{hectospec} on the MMT 6.5m telescope.
Hectospec provides simultaneous spectroscopy of up to 300 objects
across a diameter of 1$^\circ$.  This telescope and instrument
combination is ideal for studying the virial regions and outskirts of
clusters at these redshifts.

\subsubsection{Previous Spectroscopy: CIRS, HeCS, and HeCS-SZ}

The Cluster Infall Regions in SDSS (CIRS) program \citep{cirsi} studied the spectroscopic 
properties of 72 X-ray selected clusters
using data from the Fourth Data Release of SDSS \citep{dr4}.  
These galaxies are primarily in the Main Galaxy Sample and thus 
the cluster redshift surveys are
reasonably complete to $r$=17.77. 

The Hectospec Cluster Survey (HeCS) is a spectroscopic survey of 
58 galaxy clusters at moderate redshift ($z$=0.1-0.3)
with  MMT/Hectospec.  
HeCS clusters were selected based on X-ray flux. 

Because cluster properties such as 
projected velocity dispersion depend on radius, wide-field spectroscopic coverage is 
important for measuring accurate global velocity dispersions 
and virial masses \citep{biviano06}.
We used the red sequence to preselect likely
cluster members as primary targets  (targets within $\pm$0.3 mag of the red sequence).
We then filled otherwise unassigned fibers with bluer
targets \citep[][describes the details of target selection]{hecsultimate}.

HeCS-SZ includes clusters with $z<0.2$ selected based on integrated Compton parameter 
measurements from the Planck  satellite \citep{planckszcatalog} and photometry available from DR10. 
The spectroscopic observing strategy for HeCS-SZ closely matches HeCS, 
however, the color selection for red-sequence candidates is only $\pm$0.2 mag.  

Clusters from CIRS, HeCS, and HeCS-SZ in the redMaPPer catalog but outside 
the HeCS-red sample are included in the sample called HeCS-red-ext.

\subsubsection{Spectroscopy: HeCS-red}

Figure \ref{hecsredzlambda} shows the selection of the HeCS-red sample.
Although these clusters were selected based on their redMaPPer richness,
two of the newly-observed clusters (A115, A2390) 
lie above the X-ray limit of HeCS and the SZ limit of HeCS-SZ
but were not included in those surveys.

Our observing strategy closely matches HeCS and HeCS-SZ: we used SDSS 
photometry within $3\arcm$ of the redMaPPer position to identify a red sequence in 
$g-r$ color and $r$-band apparent magnitude in each cluster field.  
We use composite model magnitudes corrected for Galactic 
extinction. 
For each cluster, we then identify
a cutoff in apparent magnitude  depending on 
the redshift and richness of the target cluster.  Specifically, for each 
cluster, we choose a limiting magnitude that 
offers a good compromise of high 
completeness (sparser targets produce fewer fiber conflicts) and dense 
sampling.  The limiting apparent magnitude ranges from $r=20.0$ 
(low-redshift, rich cluster) to $r=21.2$ (higher-redshift, less rich cluster).
Targets are primarily drawn from galaxies with $g-r$ colors within 
0.2 mag of the red sequence (observed-frame colors), and we assign higher priorities to brighter
galaxies and galaxies closer to the cluster center.  This approach provides 
reasonably high sampling in the cluster cores but can lead to relatively sparse 
sampling of dense regions outside the core.  Experience from HeCS and HeCS-SZ 
indicates that two Hectospec configurations yield fairly high completeness 
for bright galaxies in cluster virial regions for
samples of 700-800 primary target galaxies. We included galaxies with slightly 
bluer colors (up to 0.4 mag bluer than the red sequence) as targets to fill any unused fibers. 
We matched all targets to redshifts from the literature as compiled 
by NED\footnote{http://ned.ipac.caltech.edu} as of 2015 September as well as to 
SDSS DR10 spectra. Most 
of the targets with existing redshifts are from SDSS, but several are from targeted 
studies of individual clusters \citep[e.g.,][for A2390 and A0115 respectively]{1996ApJ...471..694A,barrena07b}.
Targets with existing redshifts are removed from the targeting catalogs prior 
to fiber assignment.  We later supplemented the 
redshift catalogs with additional redshifts from SDSS DR13 \citep{dr13arxiv}.

Table \ref{hecsredshifts} lists \nczhecsred new redshifts measured 
with Hectospec in the fields of \nclusterhecsred clusters.  We visually inspected all spectra to confirm the reliability of 
the redshifts.  Column 5 of Table \ref{hecsredshifts} lists the cross-correlation
score $R_{XC}$ from the IRAF package {\em rvsao} \citep{km98}.  A score of 
$R_{XC}>3$ indicates a reliable redshift; some galaxies with smaller values 
of $R_{XC}$ are included when visual inspection shows multiple obvious 
absorption and/or emission lines and the spectrum suffers from contamination
(e.g., light bleeding into the spectrum from a nearby fiber containing a bright star). 
The results of visual inspection are listed as a Flag with possible values Q 
(unambiguous redshift), ? (medium-confidence redshift), and X (low-confidence redshift).
Table \ref{hecsmemredshifts} lists redshifts from SDSS 
and other literature (as compiled by NED) for galaxies classified as cluster 
members by the caustic technique (see below).  
Prior to the Hectospec observations, we measured three redshifts 
of bright galaxies with the 
FAST instrument \citep{fast} on the 
1.5-meter Tillinghast telescope at the Fred Lawrence Whipple Observatory.  
Two of these galaxies were observed with Hectospec with concordant redshifts, 
the third galaxy is at coordinates $(\alpha,\delta)$=(23:26:26.15,+29:21:52.67) 
and has heliocentric redshift $cz_\odot=(67891.9\pm75)~\kms $.  This galaxy is a 
member of RMJ2326. 

\begin{figure*} 
\plotone{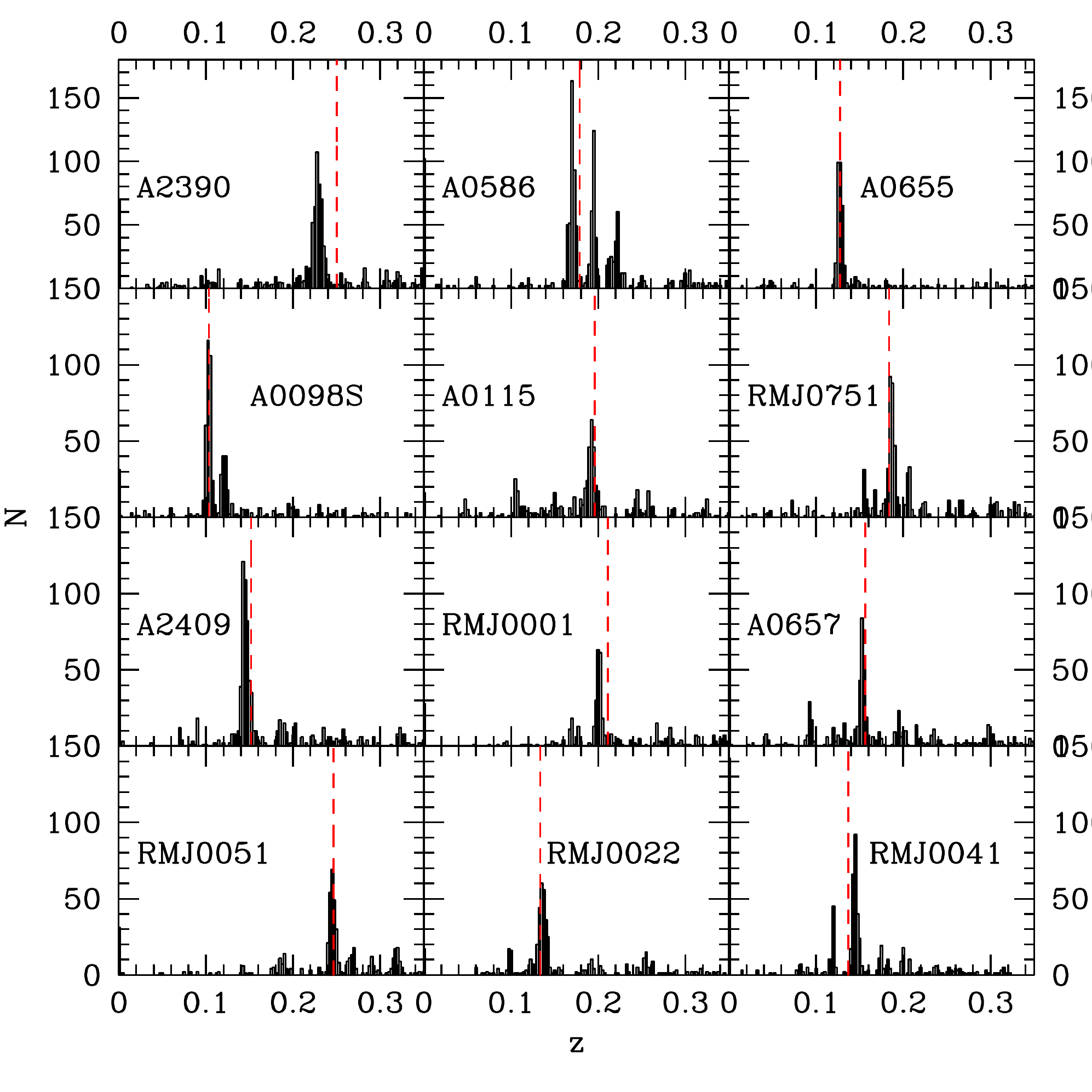}
\caption{\label{hecsredzhist1} Redshift histograms for galaxies within 0.5 degrees of the cluster centers for the redMaPPer sample.  
Bins have width $\Delta z = 0.0025$. Red dashed lines indicate the photometric redshifts from redMaPPer.  
}
\end{figure*}

\begin{figure*} 
\plotone{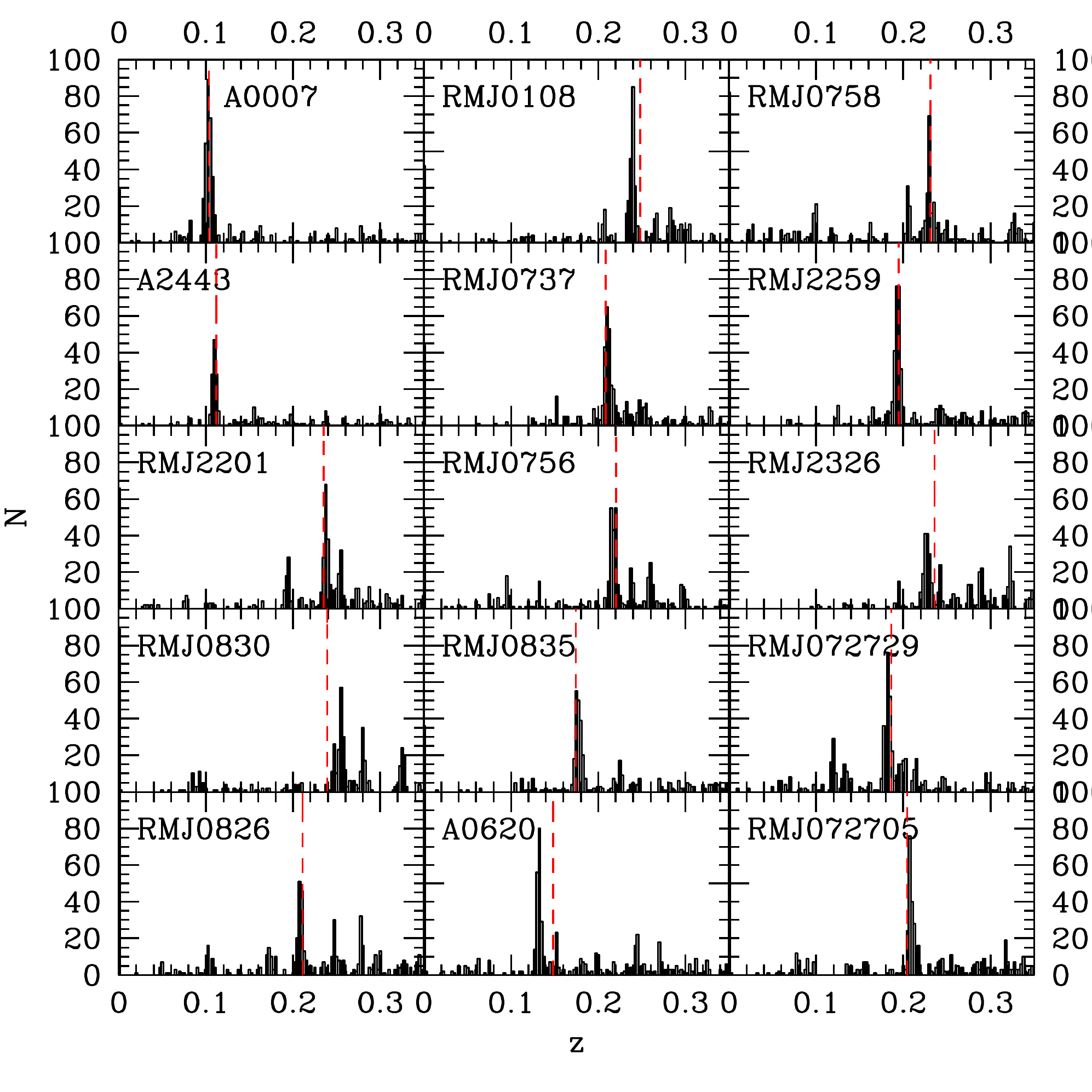}
\caption{\label{hecsredzhist2} Redshift histograms for galaxies within 0.5 degree of the cluster centers for the redMaPPer sample.  
Bins have width $\Delta z = 0.0025$. Red dashed lines indicate the photometric redshifts from redMaPPer.
}
\end{figure*}

Figures \ref{hecsredzhist1}-\ref{hecsredzhist2} show redshift histograms of the HeCS-red clusters. 
Each of the clusters in the sample 
shows a prominent peak in the redshift histogram close to the location of the 
redMaPPer photometric redshift.  Several clusters (A0586, A0098S, RMJ0751, RMJ2201, RMJ2326, RMJ0830, RMJ0826)
show additional peaks  that could contaminate the richness estimates.

Our primary sample is the 90\%-complete HeCS-red, selected exclusively 
from the redMaPPer catalog.  HeCS-red-ext is an extended sample that 
includes clusters in the redMaPPer catalog  with existing spectroscopy from 
CIRS, HeCS, or HeCS-SZ.  HeCS-red-ext also 
includes two high-richness clusters (RMJ0230 and A2355) with new 
MMT/Hectospec data listed in Table 1. These two clusters lie outside the declination 
range of HeCS-red.  
Figure \ref{hecsredext} shows redshift histograms and phase space diagrams of these two clusters. 
In total, HeCS-red-ext includes 121 clusters in a heterogeneously selected sample.

\begin{figure*} 
\plotone{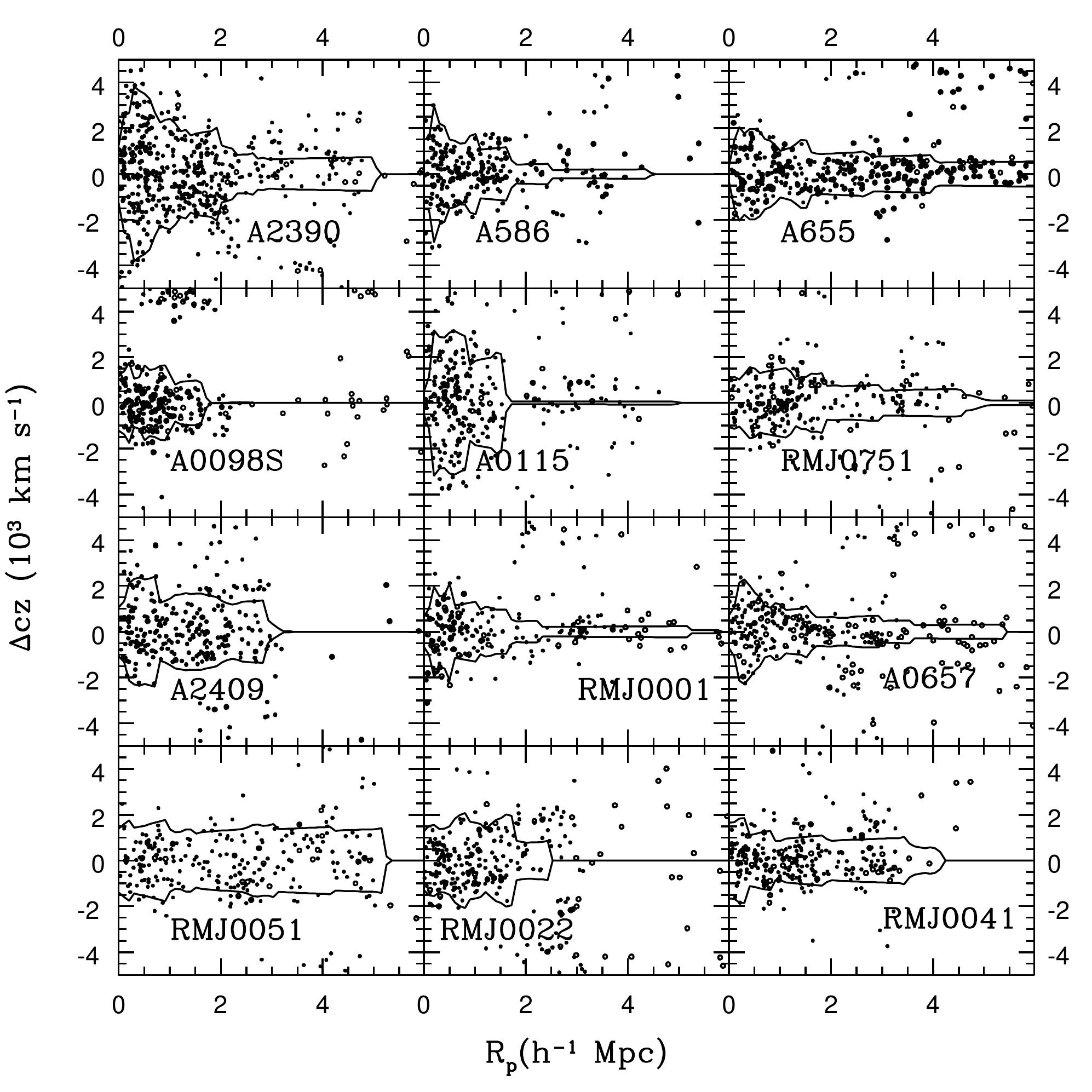}
\caption{\label{hecsredc1} {Redshift (rest-frame clustrocentric velocity) versus projected radius for galaxies around
HeCS-red clusters.  The caustic
pattern is evident as the trumpet-shaped regions with high density.
The solid lines indicate our estimate of the location of the caustics
in each cluster.  Clusters are ordered left-to-right and top-to-bottom
by decreasing richness parameter $\lambda$.  }}
\end{figure*}

\begin{figure*} 
\plotone{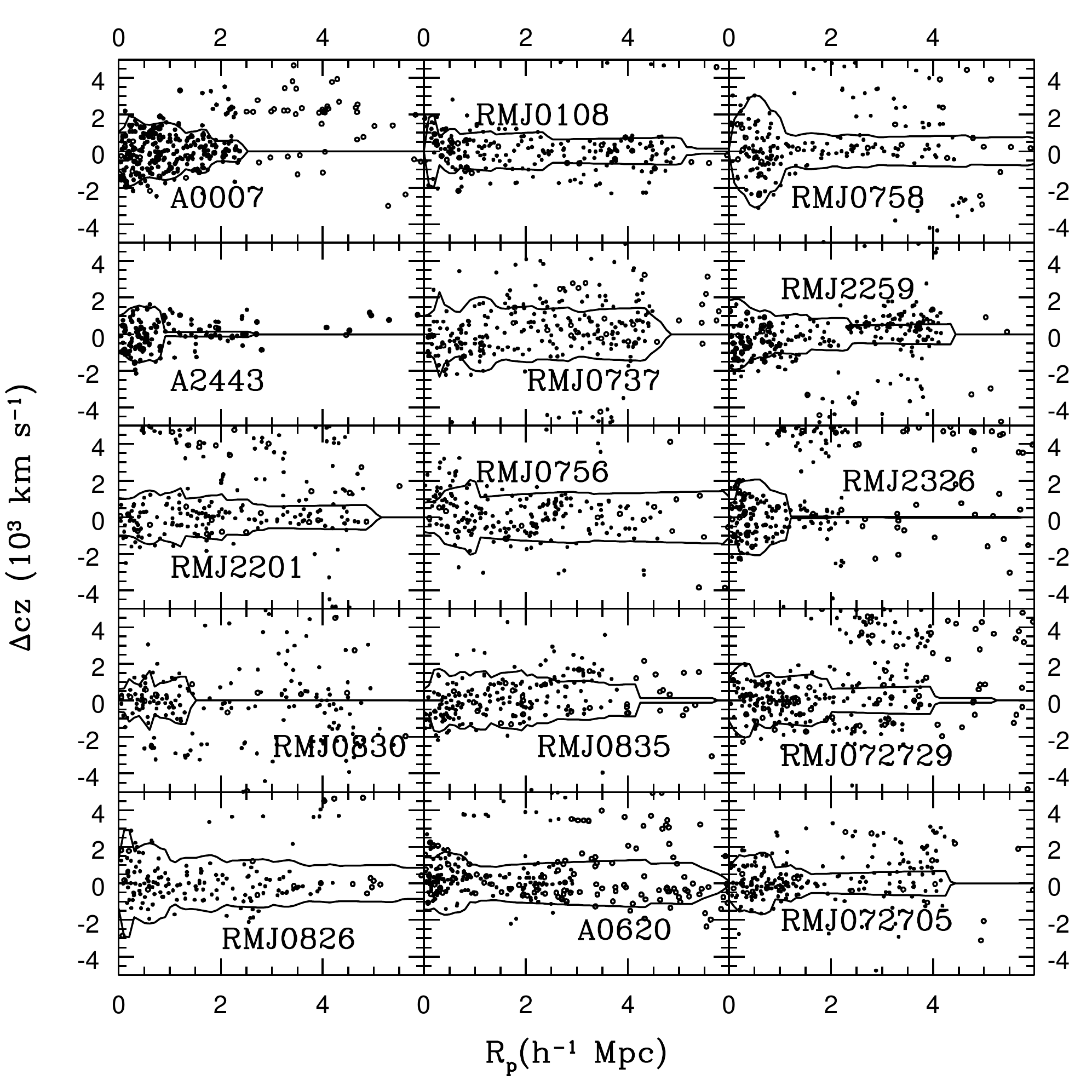}
\caption{\label{hecsredc2} {Same as Figure \ref{hecsredc1}.   }}
\end{figure*}

\begin{figure} 
\plotone{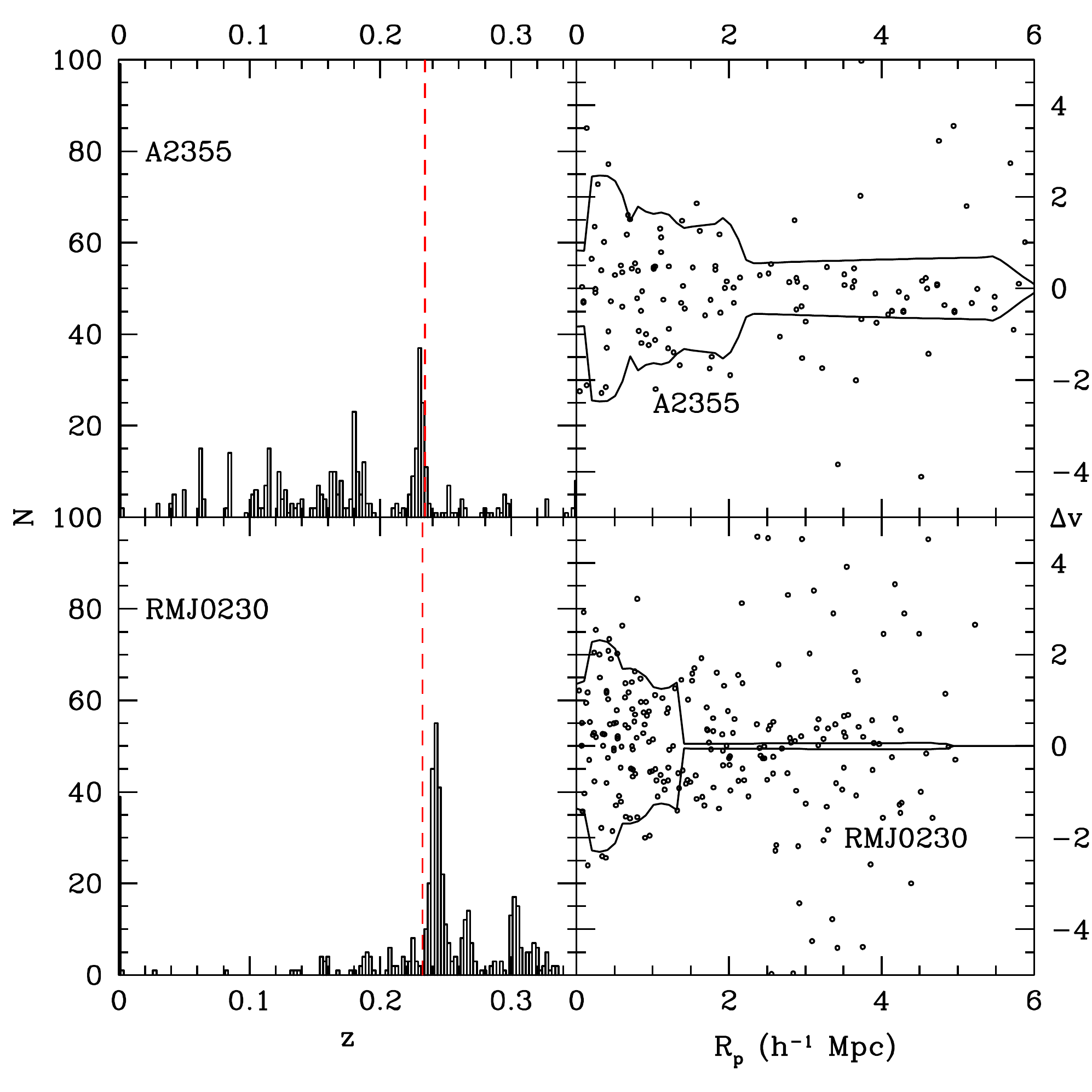}
\caption{\label{hecsredext} Similar to Figures \ref{hecsredzhist1} and \ref{hecsredc1} for the 
HeCS-red-ext clusters A2355 and RMJ0230 with redshifts in Table \ref{hecsredshifts}. Note that the target catalog 
for A2355 did not use a color cut.
}
\end{figure}

\begin{table}[th] \footnotesize
\begin{center}
\caption{\label{hecsredshifts} \sc HeCS-red Redshifts from MMT/Hectospec}
\begin{tabular}{ccrrrcc}
\tableline
\tableline
\tablewidth{0pt}
\multicolumn{2}{c}{Coordinates (J2000)} & $cz_\odot$ & $\sigma_{cz}$  & $R_{XC}$ & Flag & Member \\ 
 RA  & DEC  &  km/s &km/s & &  & \\ 
\tableline
0:00:03.11  &   12:05:18.63  & 60460.6 & 26.4  & 16.75 & Q  & 1 \\ 
0:00:03.17  &   11:57:28.94  & 80131.3 & 67.9  & 6.07 & Q  & 0 \\ 
0:00:06.46  &   11:58:17.04  & 42771.2 & 49.5  & 6.38 & Q  & 0 \\ 
0:00:07.50  &   11:59:05.43  & 60979.7 & 37.7  & 10.55 & Q  & 0 \\ 
0:00:09.45  &   11:54:14.53  & 38946.6 & 13.3  & 9.07 & Q &  0 \\ 
\tableline
\end{tabular}
\tablecomments{Table \ref{hecsredshifts} is published in its entirety in the electronic edition of the Journal. A portion is shown here for guidance regarding its form and content.}
\end{center}
\end{table}

\begin{table}[th] \footnotesize
\begin{center}
\caption{\label{hecsmemredshifts} \sc HeCS-red Members from Literature Redshifts }
\begin{tabular}{ccrrc}
\tableline
\tableline
\tablewidth{0pt}
\multicolumn{2}{c}{Coordinates (J2000)} & $cz_\odot$ & $\sigma_{cz}$  & Ref. \\
 RA  & DEC  &  km/s &km/s & \\ 
\tableline
0:01:39.655   &  12:03:11.231  &  61067 & 13  &   1        \\
0:01:39.919   &  12:04:02.063  &  61348 & 18  &   1        \\
0:01:40.277   &  11:55:32.915  &  59691 & 06  &   1        \\
0:01:43.944   &  12:03:54.143  &  60366 & 20  &   1        \\
0:01:44.863   &  12:03:56.089  &  60443 & 09  &   1        \\
\tableline
\end{tabular}
\tablecomments{Table \ref{hecsmemredshifts} is published in its entirety in the electronic edition of the Journal. A portion is shown here for guidance regarding its form and content.}
\tablecomments{References: [1] SDSS, [2] NED. }
\end{center}
\end{table}

\subsection{Cluster Redshifts, Velocity Dispersions, and Masses}

The caustic technique \citep{dg97,diaferio1999,serra11} isolates cluster members from foreground and background galaxies 
in phase space.  After smoothing the galaxy distribution in the redshift diagram, the infall regions of 
clusters produce well-defined envelopes containing the vast majority of cluster 
members.  In numerical simulations, 96\% of cluster members within $r_{200}$ lie 
inside the caustic envelope, and 
only 2\% of galaxies inside the caustic envelope are actually interlopers. Within the larger
radius $3r_{200}$ , where the caustic technique is the only usable method, the completeness is 95\% and the interloper fraction is 8\% \citep{serra12}.
The edges of this distribution are called caustics and they are related 
to the escape velocity profile of the cluster \citep[see][for reviews]{diaferio09,serra11}.  
The escape velocity profile is the basis for a mass profile that can extend 
into the infall region where the galaxies are gravitationally bound but not virialized.
Caustic mass estimates generally agree with
estimates from X-ray observations and gravitational lensing
\citep[e.g.,][and references therein]{cairnsi,bg03,diaferio05,cirsi,cirsmf,hecslens}.

Figures \ref{hecsredc1}-\ref{hecsredc2} show 
the phase space 
diagrams of the HeCS-red clusters not already published in HeCS or 
HeCS-SZ.  All clusters display  infall 
patterns: the caustics are shown on the figures.
Clusters are ordered by decreasing richness parameter $\lambda$.
There is a general trend of decreasing central velocity dispersion 
with decreasing richness.  Figure \ref{hecsredext} shows the 
phase space diagrams for the HeCS-red-ext clusters included
in Table 1. 

We apply the prescription of \citet{danese} to determine the mean
redshift $cz_\odot$ and projected velocity dispersion $\sigma_p$ of
each cluster from all galaxies within the caustics.  We calculate
$\sigma_p$ using only the cluster members projected within $r_{200}$
estimated from the caustic mass profile.  
Note that our measured velocity dispersions use the caustic technique 
only to define membership and the limiting radius $r_{200}$.  Independent
of its performance as a mass estimator, the caustic technique is a highly 
efficient membership selection algorithm, especially at the relatively 
small radii we focus on here \citep{serra12}.  Table \ref{hecsredtab} 
lists the central cluster redshifts, velocity dispersions inside $r_{200}$,
and $M_{200}$ from the caustic mass profile. The eighth column of 
Table \ref{hecsredtab} indicates whether the cluster is part of the CIRS, 
HeCS, HeCS-SZ, or HeCS-red sample.

\subsection{Comparison of Spectroscopic Redshifts to redMaPPer Photometric Redshifts}

The photometric redshifts in the redMaPPer catalog are usually close to the central 
redshifts we obtain in our hierarchical clustering analysis of the cluster 
redshifts (see D99 for details).  However, for about half of the clusters,
our central redshifts differ by more than a percent from the redMaPPer photometric redshifts.  

\begin{figure} 
\plotone{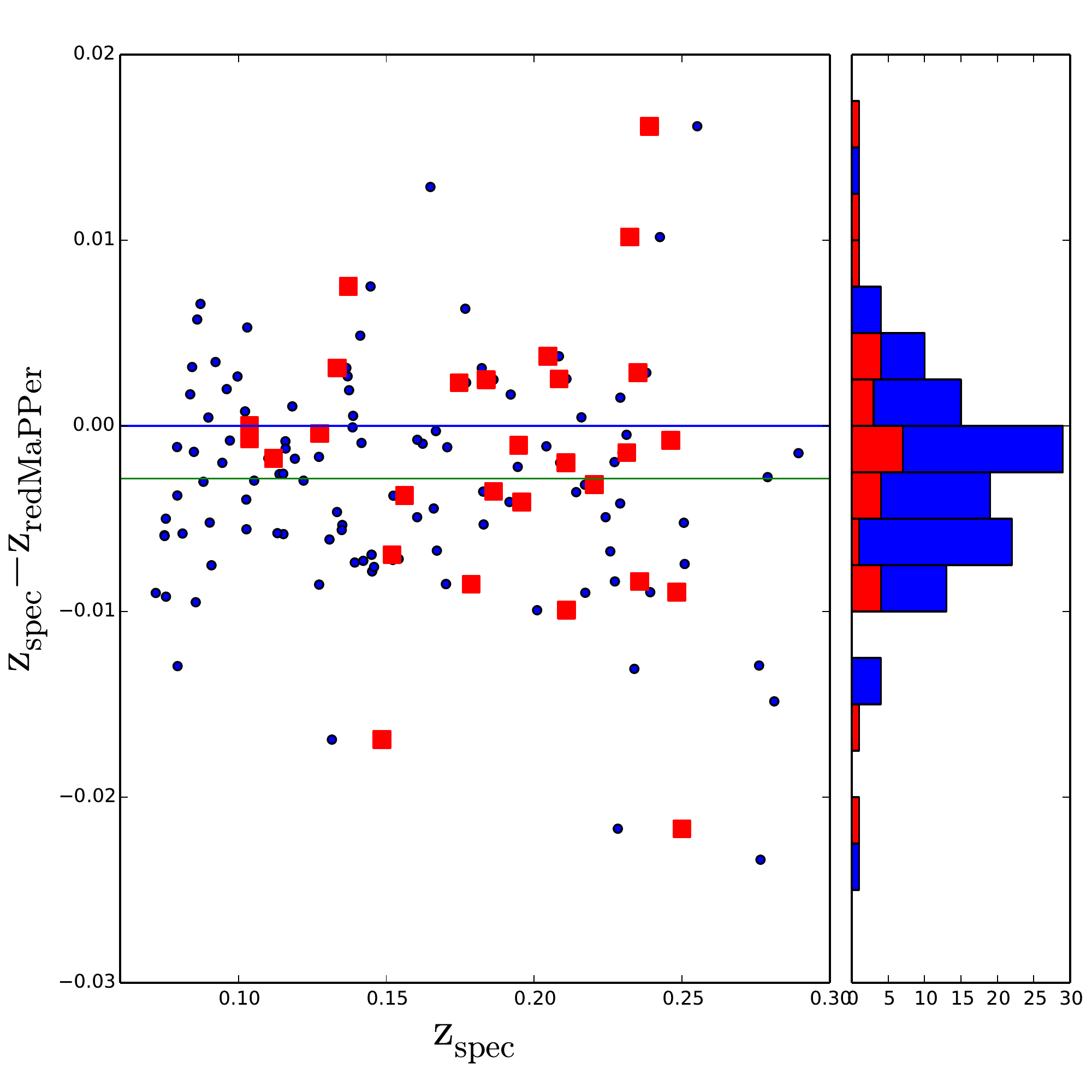}
\caption{\label{hecsredzphotz} Spectroscopic redshifts versus photometric redshifts from the redMaPPer catalog.
Red squares and blue circles show clusters from HeCS-red and HeCS-red-ext respectively.  
The blue horizontal lines shows zero offset. The green horizontal line shows the mean offset of the HeCS-red-ext sample.
}
\end{figure}

Figure \ref{hecsredzphotz} compares photometric redshift estimates from the 
redMaPPer catalog to the spectroscopic redshifts.  For the HeCS-red sample, 
the mean offset is $z_{spec}-z_{RM}=-0.0019\pm0.0014$.  For HeCS-red-ext, 
the mean offset is $z_{spec}-z_{RM}=-0.0028\pm0.0005$ (green line in 
Figure \ref{hecsredzphotz}), suggesting a bias in the photometric redshifts 
at $5.5\sigma$ confidence level.  Comparisons of redMaPPer redshifts to 
BCG redshifts in DR8 showed no such offset \citep{rykoff14}.  Studies of BCG 
redshifts relative to their clusters shows that, while most BCGs are located 
close to the mean redshift of their clusters, some BCGs have significant 
velocity offsets \citep[e.g.,][]{lauer14}.

The redshift bias observed for the HeCS-red-ext sample could be produced 
by an excess of background contamination over foreground contamination.  
The photometric redshift window surrounding the cluster contains a larger volume
at redshifts above than the cluster redshift than the volume below than the 
cluster redshift.

Spectroscopic targets for A2355 were selected by apparent magnitude only (no color cut).  
Figure \ref{hecsredext} shows that this cluster is not as well sampled 
as clusters of similar richness where target selection prioritized red-sequence galaxies. 
This difference highlights the greater efficiency of red-sequence target selection for 
identifying cluster members. 
 
\begin{table*}[th] \footnotesize
\begin{center}
\caption{\label{hecsredtab} \sc  Dynamical Mass and Richness Estimates}
\begin{tabular}{lcccccccc}
\tableline
\tableline
\tablewidth{0pt}
Cluster & $\alpha$ & $\delta$ & $z_\odot$    & $\sigma_p$ & $M_{200,c}$ & $\lambda$  & Spectra & redMaPPer ID  \\ 
 & $\deg$ & $\deg$ &    & $\kms$ & $10^{14} M_\odot$ &  &  \\ 
\tableline
 A2390 & 328.39839 & 17.69735 & 0.2283  & $ 1278 ^{+64} _{-55}$ & $ 12.10 {\pm 0.62} $ & $ 149.3 \pm{5.5} $ & HeCS-red & RMJ215336.8+174143.7    \\ 
 A0586 & 113.09431 & 31.62882 & 0.1702  & $ 797 ^{+55} _{-45}$ & $ 4.34 {\pm 0.18} $ & $ 145.1 \pm{5.5} $ & HeCS-SZ & RMJ073220.3+313800.7    \\ 
 A0655 & 126.34018 & 47.15855 & 0.1271  & $ 777 ^{+58} _{-47}$ & $ 3.33 {\pm 0.16} $ & $ 130.7 \pm{5.4} $ & HeCS & RMJ082529.1+470800.9    \\ 
 A0098S & 11.63442 & 20.47665 & 0.1038  & $ 624 ^{+47} _{-38}$ & $ 2.42 {\pm 0.11} $ & $ 127.1 \pm{3.3} $ & HeCS-SZ & RMJ004629.3+202804.8    \\ 
 A0115 & 14.00240 & 26.33962 & 0.1916  & $ 1176 ^{+70} _{-59}$ & $ 10.60 {\pm 0.16} $ & $ 111.5 \pm{3.7} $ & HeCS-red & RMJ005600.3+262032.3    \\ 
 RMJ075100.8+173753.8 & 117.81519 & 17.65724 & 0.1863  & $ 637 ^{+68} _{-51}$ & $ 1.72 {\pm 0.07} $ & $ 110.5 \pm{6.1} $ & HeCS-red & RMJ075100.8+173753.8    \\ 
 A2409 & 330.22102 & 20.96189 & 0.1450  & $ 1038 ^{+79} _{-64}$ & $ 5.30 {\pm 0.38} $ & $ 92.8 \pm{4.6} $ & HeCS-SZ & RMJ220052.6+205809.3    \\ 
 RMJ000158.5+120358.0 & 00.50730 & 12.07572 & 0.2010  & $ 647 ^{+67} _{-51}$ & $ 2.38 {\pm 0.22} $ & $ 84.9 \pm{3.5} $ & HeCS-red & RMJ000158.5+120358.0    \\ 
 A0657 & 125.83242 & 15.95854 & 0.1524  & $ 777 ^{+62} _{-50}$ & $ 3.01 {\pm 0.39} $ & $ 81.6 \pm{3.6} $ & HeCS-red & RMJ082319.3+155745.8    \\ 
 RMJ005105.2+261716.7 & 12.76264 & 26.30037 & 0.2454  & $ 661 ^{+66} _{-51}$ & $ 2.97 {\pm 0.05} $ & $ 81.6 \pm{4.0} $ & HeCS-red & RMJ005105.2+261716.7    \\ 
 RMJ002224.7+231733.0 & 05.59923 & 23.29194 & 0.1365  & $ 710 ^{+49} _{-41}$ & $ 3.51 {\pm 0.21} $ & $ 80.8 \pm{3.3} $ & HeCS-red & RMJ002224.7+231733.0    \\ 
 RMJ004118.5+252609.1 & 10.31908 & 25.43062 & 0.1447  & $ 652 ^{+64} _{-49}$ & $ 2.24 {\pm 0.06} $ & $ 80.0 \pm{3.5} $ & HeCS-red & RMJ004118.5+252609.1    \\ 
 A0007 & 02.94185 & 32.42523 & 0.1030  & $ 816 ^{+55} _{-46}$ & $ 3.03 {\pm 1.28} $ & $ 79.3 \pm{4.1} $ & HeCS-SZ & RMJ001145.3+322456.4    \\ 
 RMJ010819.0+275802.1 & 17.07981 & 27.96756 & 0.2393  & $ 609 ^{+81} _{-58}$ & $ 1.55 {\pm 0.02} $ & $ 78.1 \pm{3.7} $ & HeCS-red & RMJ010819.0+275802.1    \\ 
 RMJ075822.7+264120.6 & 119.62574 & 26.68098 & 0.2298  & $ 986 ^{+80} _{-64}$ & $ 7.50 {\pm 0.21} $ & $ 77.9 \pm{4.0} $ & HeCS-red & RMJ075822.7+264120.6    \\ 
 A2443 & 336.50488 & 17.37134 & 0.1102  & $ 652 ^{+58} _{-46}$ & $ 1.93 {\pm 0.64} $ & $ 77.6 \pm{2.8} $ & HeCS-SZ & RMJ222607.9+172123.4    \\ 
 RMJ073720.9+351741.7 & 114.33470 & 35.28469 & 0.2109  & $ 602 ^{+63} _{-48}$ & $ 2.80 {\pm 0.25} $ & $ 76.6 \pm{4.3} $ & HeCS-red & RMJ073720.9+351741.7   \\ 
 RMJ225946.5+310223.9 & 344.95589 & 31.03735 & 0.1937  & $ 661 ^{+61} _{-48}$ & $ 2.60 {\pm 0.03} $ & $ 75.2 \pm{3.5} $ & HeCS-red & RMJ225946.5+310223.9   \\ 
 RMJ220107.7+111805.2 & 330.28072 & 11.29804 & 0.2379  & $ 560 ^{+77} _{-54}$ & $ 1.41 {\pm 0.15} $ & $ 74.1 \pm{4.4} $ & HeCS-red & RMJ220107.7+111805.2   \\ 
 RMJ075655.8+383933.2 & 119.26340 & 38.68267 & 0.2172  & $ 683 ^{+87} _{-63}$ & $ 1.99 {\pm 0.02} $ & $ 72.7 \pm{3.7} $ & HeCS-red & RMJ075655.8+383933.2   \\ 
 RMJ083056.4+322412.2 & 127.69104 & 32.45600 & 0.2551  & $ 482 ^{+75} _{-51}$ & $ 0.88 {\pm 0.12} $ & $ 67.6 \pm{4.1} $ & HeCS-red & RMJ083056.4+322412.2   \\ 
 RMJ083513.0+204654.9 & 128.76012 & 20.78112 & 0.1770  & $ 611 ^{+64} _{-49}$ & $ 2.22 {\pm 0.02} $ & $ 67.5 \pm{4.2} $ & HeCS-red & RMJ083513.0+204654.9   \\ 
 RMJ232626.2+292152.7 & 351.60948 & 29.35085 & 0.2273  & $ 938 ^{+95} _{-73}$ & $ 3.63 {\pm 0.14} $ & $ 68.9 \pm{5.6} $ & HeCS-red & RMJ232626.2+292152.7   \\ 
 RMJ072729.3+422756.1 & 111.88434 & 42.51032 & 0.1828  & $ 702 ^{+60} _{-48}$ & $ 3.04 {\pm 0.14} $ & $ 66.9 \pm{3.8} $ & HeCS-red & RMJ072729.3+422756.1   \\ 
 RMJ082657.6+310804.9 & 126.72990 & 31.14427 & 0.2088  & $ 890 ^{+86} _{-67}$ & $ 6.08 {\pm 0.32} $ & $ 65.7 \pm{3.2} $ & HeCS-red & RMJ082657.6+310804.9   \\ 
 A0620 & 121.43548 & 45.67952 & 0.1316  & $ 702 ^{+59} _{-47}$ & $ 2.57 {\pm 0.07} $ & $ 64.3 \pm{2.5} $ & HeCS-red & RMJ080543.3+454058.9   \\ 
 RMJ072705.2+384613.4 & 111.77754 & 38.80076 & 0.2084  & $ 614 ^{+59} _{-46}$ & $ 2.35 {\pm 0.04} $ & $ 64.0 \pm{3.8} $ & HeCS-red & RMJ072705.2+384613.4   \\ 
\tableline
 RMJ023054.9+024719.6 & 37.72578 & 02.78778 & 0.2425  & $ 878 ^{+77} _{-61}$ & $ 4.34 {\pm 0.92} $ & $ 90.7 \pm{4.1} $ & HeCS-red-ext & RMJ023054.9+024719.6    \\ 
 A2355    & 323.81759 & 01.39962 & 0.2306  & $ 911^{+125} _{-88}$ & $ 5.12 {\pm 0.73} $ & $109.4\pm{4.8}$ & HeCS-red-ext & RMJ213518.8+012527.0     \\ 
\tableline
\end{tabular}
\end{center}
\tablecomments{Redshift $z$ and velocity dispersion $\sigma_p$ are computed for galaxies defined as members using the caustics.   }
\end{table*}

\section{Cluster Scaling Relations}

\subsection{Bayesian parameter estimation}\label{sec:bayesParam}

We determine the scaling relations between redMaPPer cluster richness
$\lambda$ and spectroscopic properties for the HeCS-red sample
using a Bayesian approach similar to HeCS-SZ.  
A number of unknown hidden variables produces a scatter
in the linear correlation $Y=a+bX$. We model this scatter with a
single parameter, the intrinsic dispersion $\sigma_{\rm int}$.
Therefore, given a measure $X_i$ with uncertainty $\sigma_{X_i}$, the probability
of measuring $Y_i$ with uncertainty $\sigma_{Y_i}$
is $p(Y_i,\sigma_{Y_i}\vert \theta,X_i,\sigma_{X_i})$, where $\theta=\{a,b,\sigma_{\rm int}\}$.
We assume the Gaussian likelihood
\begin{equation}
p(D\vert\theta,M) = \prod_i {1\over (2\pi \sigma_i^2)^{1/2}} \exp\left[-(Y_i-a-bX_i)^2 \over 2\sigma_i^2\right]  
\end{equation}
where $M$ is the model with parameter set $\theta$, $D$ is the data, and 
\begin{equation}
\sigma_i^2=\sigma_{\rm int}^2 + \sigma_{Y_i}^2 + b^2 \sigma_{X_i}^2 \; .
\end{equation}

We assume independent flat priors for both $a$ and $b$.
For the intrinsic dispersion $\sigma_{\rm int}$, which is
positive defined, we assume
\begin{equation}
p(\sigma_{\rm int}\vert M) = {\mu^r\over \Gamma(r)}  x^{r-1} \exp(-\mu x) 
\end{equation}
where $x=1/\sigma_{\rm int}^2$, and $\Gamma(r)$ is the usual gamma function.
This PDF describes a variate with mean $ r/\mu$, and variance $ r/\mu^2$.
We set $r=\mu=10^{-5}$ which guarantees an almost flat prior.

We estimate the parameter PDF $p(\theta\vert D, M)$ via Markov Chain Monte Carlo (MCMC)
sampling with the code APEMoST  \citep{buchner11, gruberbauer09}.
We obtain a fairly complete sampling with $2\times 10^6$ MCMC iterations.
The boundaries of the parameter space were set to
$[-100,100]$ for $a$ and $b$, and $[0.01,100]$ for $\sigma_{\rm int}$.
As the three best-fit parameters $a$, $b$, and $\sigma_{\rm int}$ of the Bayesian 
analysis, we adopt the medians derived from the posterior PDF $p(\theta\vert D,M)$. 
Likewise, we adopt the boundaries of the 68\% credible intervals around the 
medians as the uncertainties on these best-fit parameters.
Table \ref{hecsredfits} lists the best-fit parameters.

\subsection{Scaling of Richness with Velocity Dispersion and Dynamical Mass}

\begin{figure*} 
\epsscale{1.0}
\plotone{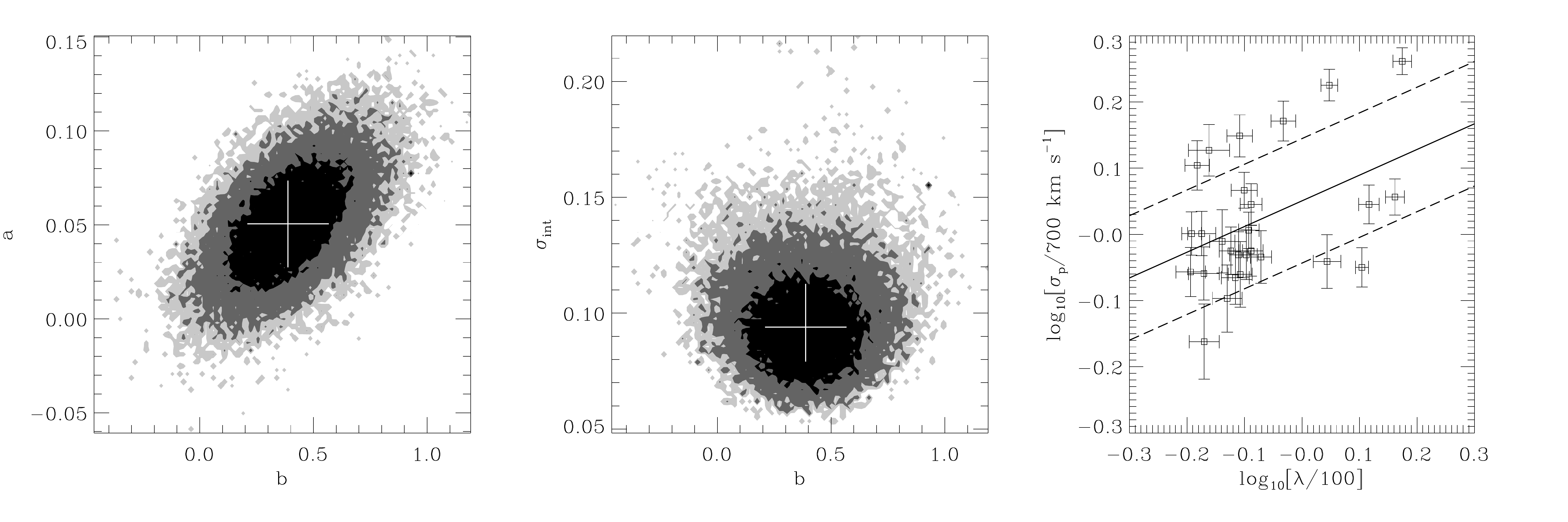}
\caption{\label{sigmapvsrichred} (Left panel) Marginalized probability distribution function of the intercept and slope 
of the relation $P(\sigma_p|\lambda) = a + b~ {\rm log}(\lambda)$ from MCMC analysis.  The shaded areas with decreasing darkness show the 68.3\%, 95.4\%, and 99.7\% marginalized credible intervals.  The white cross shows the 68.3\% marginalized credible interval of each parameter.
(Middle) Similar to the left panel for the intrinsic scatter $\sigma_{\rm int}$ and intercept $b$ of the scaling relation.
(Right) Scaling relation between projected velocity dispersion $\sigma_p$ and 
the richness proxy $\lambda$ for clusters in the redMaPPer-complete sample (HeCS-red). 
Dashed lines indicate the intrinsic scatter for individual clusters. 
}
\end{figure*}

\begin{figure} 
\epsscale{1.0}
\plotone{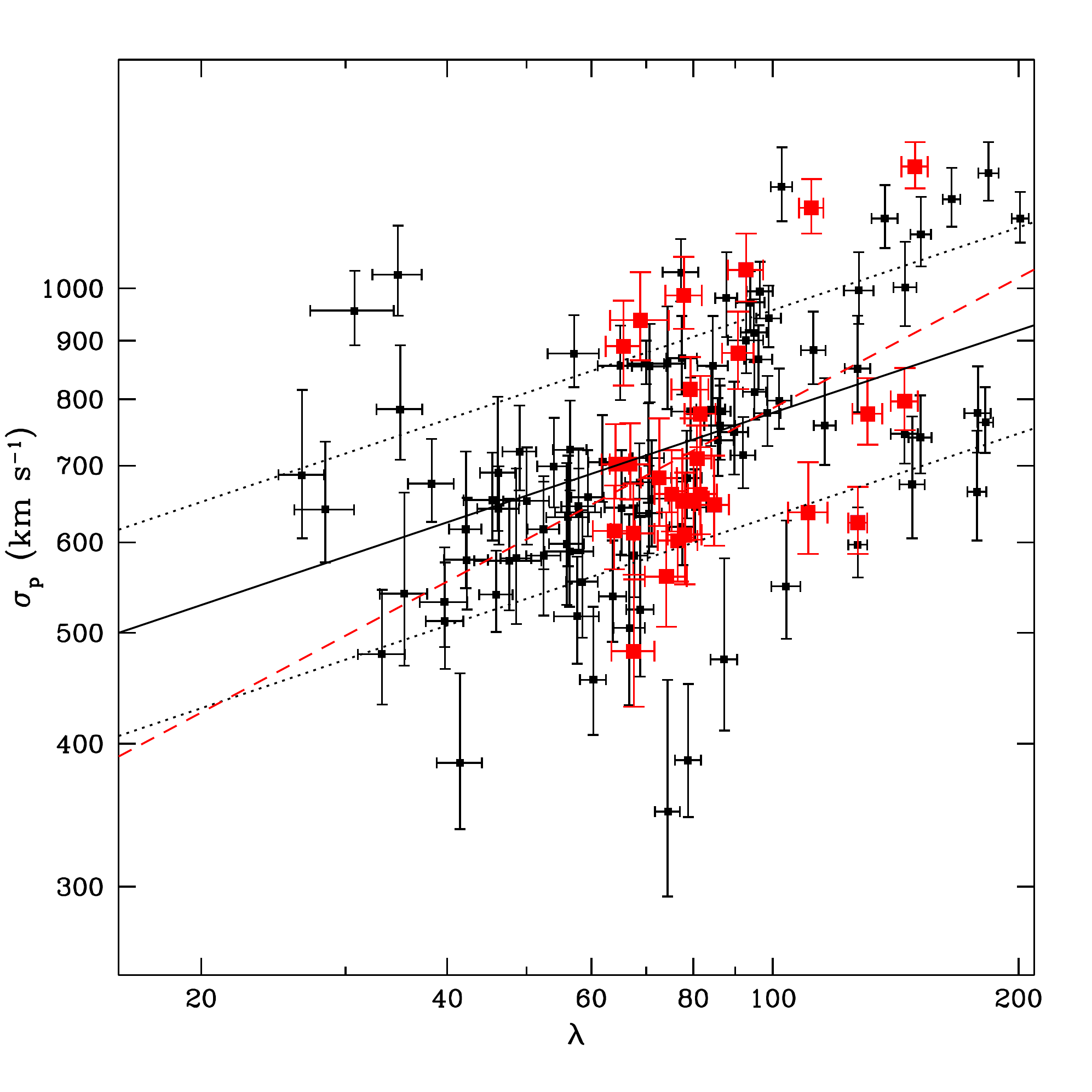}
\caption{\label{sigmapvsrich} Scaling relation between projected velocity dispersion $\sigma_p$ and 
the richness proxy $\lambda$ for clusters in the redMaPPer-complete sample  
supplemented with clusters from other surveys (Sample 2). 
Red and black points show clusters from HeCS-red and HeCS-red-ext respectively.
The thick solid line shows the best-fit relation of 
$P(\sigma_p|\lambda)$ for the HeCS-red-ext  sample with the intrinsic scatter shown as dotted lines.  
For reference, the dashed line shows the best-fit relation for the HeCS-red sample.
}
\end{figure}

Figure \ref{sigmapvsrichred} shows the best-fit relation for $P(\sigma_p|\lambda)$, 
the expected velocity dispersion at fixed richness proxy $\lambda$.  
Substantial scatter is present at fixed $\lambda$; the range of measured velocity 
dispersion at fixed $\lambda$ is a factor of two.
Figure \ref{sigmapvsrich} shows $\sigma_p$ versus $\lambda$ for the enlarged 
HeCS-red-ext sample.  The best-fit relation for the HeCS-red-ext sample agrees with 
the best-fit relation for the HeCS-red sample, indicating that sample selection does 
bias the inferred parameters of the scaling relation, at least to the accuracy of the 
samples presented here.  However, the enlarged sample contains more outliers, 
and the range of measured velocity dispersion at fixed $\lambda$ increases to a 
factor of three.

\begin{figure} 
\epsscale{1.0}
\plotone{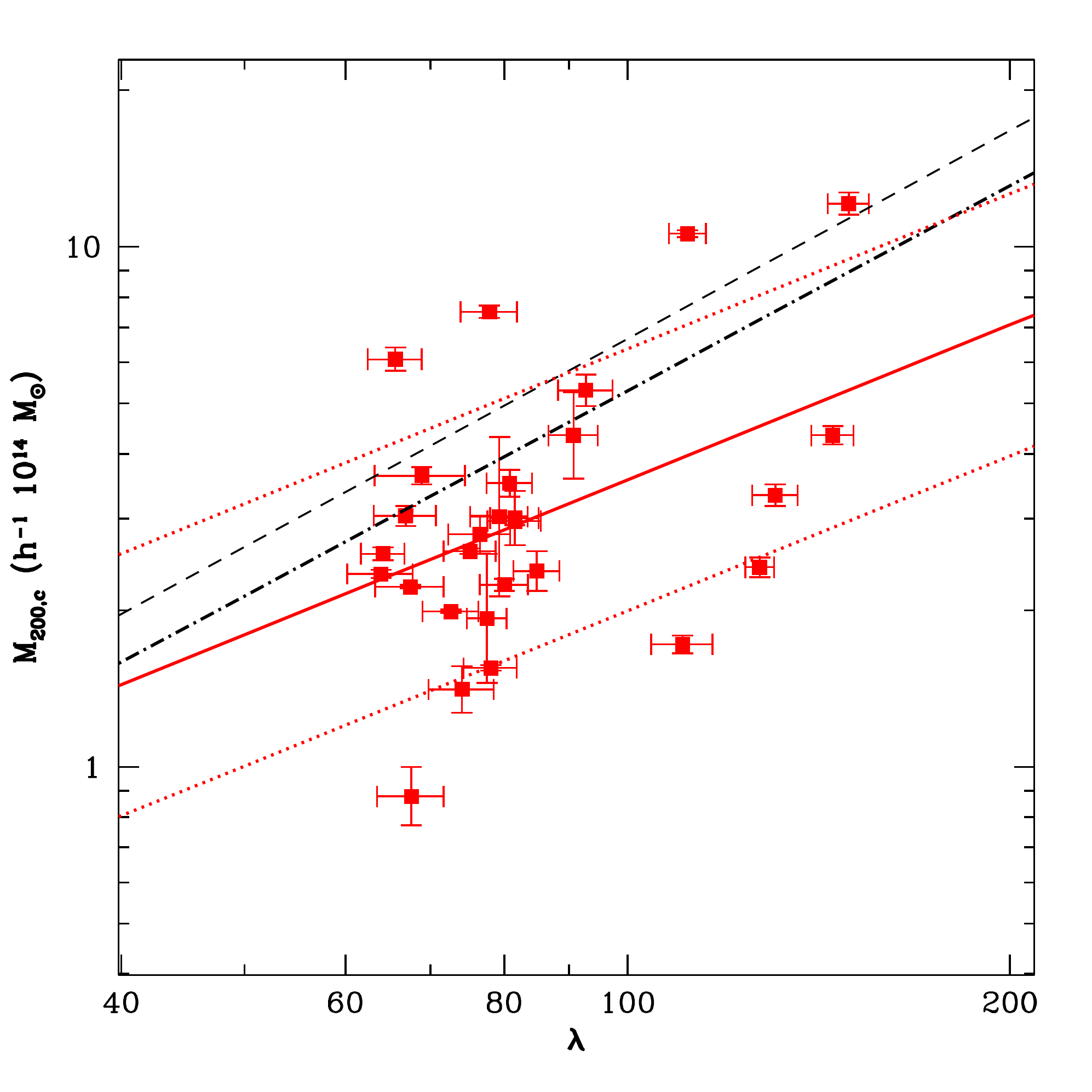}
\caption{\label{m200vslambdared} Similar to Figure \ref{sigmapvsrichred} for the scaling relation between caustic mass $M_{200}$  
and richness proxy $\lambda$ for the HeCS-red sample.
The red solid and dotted lines show the best-fit relation for HeCS-red-ext 
and the intrinsic scatter for individual clusters.  
The black dash-dotted line shows the mass-$\lambda$ scaling relation 
from a kinematic analysis of redMapper clusters using sparsely sampled spectroscopy \citep{farahi16}.  
The dashed line shows the mass-$\lambda$ scaling relation from a weak lensing analysis of stacked redMapper clusters \citep{simet17a}.
}
\end{figure}

\begin{figure} 
\epsscale{1.0}
\plotone{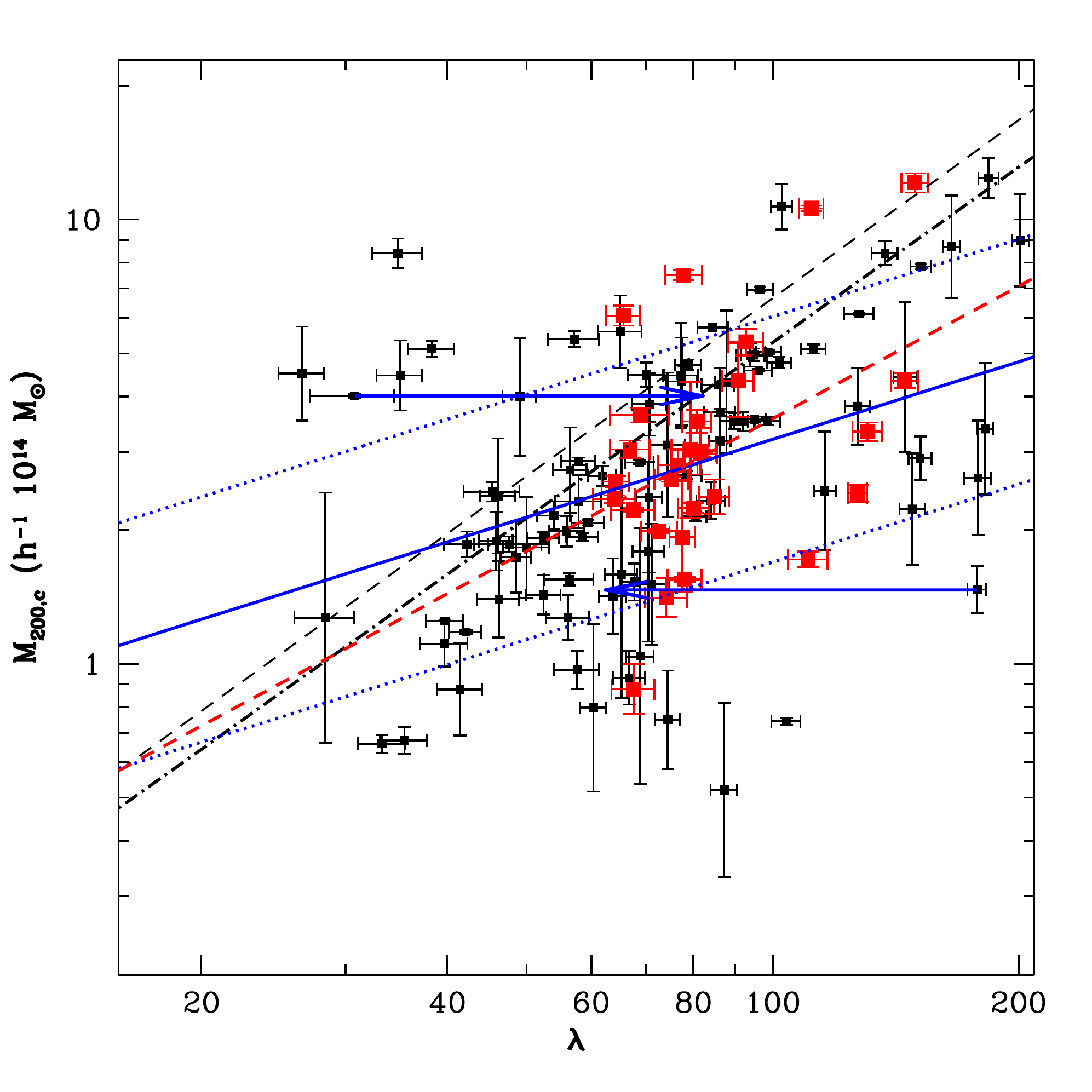}
\caption{\label{m200vslambda} Similar to Figure \ref{sigmapvsrich} for the scaling relation between caustic mass $M_{200}$  
and richness proxy $\lambda$.
Red and black points are clusters from HeCS-red and HeCS-red-ext respectively.
The blue solid and dotted lines show the best-fit relation for HeCS-red-ext 
and the intrinsic scatter for individual clusters.  For comparison, the red dashed line 
shows the best-fit relation for the HeCS-red sample.
The black dash-dotted line and dashed line show the mass-$\lambda$ scaling relations 
from \citet{farahi16} and \citet{simet17a} respectively. 
Arrows show spectroscopic richness estimates for A0963 and MS0906.  The richness of A0963 was underestimated due to problems with photometry in the redMaPPer analysis and the richness of MS0906 is overestimated due to a superposition with A0750. 
}
\end{figure}

Figure \ref{m200vslambdared} shows the best-fit relation $P(M_{200}|\lambda)$, the 
caustic mass $M_{200}$ obtained at fixed $\lambda$, for the HeCS-red sample.  
Figure \ref{m200vslambda} shows the best-fit relation for the HeCS-red-ext sample.  The intrinsic  scatter in 
this relation is about 90\% for both the HeCS-red and HeCS-red-ext samples (Table \ref{hecsredfits}).
Similar to the $\sigma_p-\lambda$ scaling relation, there is no evidence of 
bias in the inferred parameters for the HeCS-red-ext sample due to the heterogeneous 
sample selection.
The range of $M_{200}$ at fixed $\lambda$ is about a factor of 10 (20) for 
the HeCS-red (HeCS-red-ext) sample, consistent with scaling the range of measured 
velocity dispersions at fixed $\lambda$ by a virial scaling relation 
$M_{200}\propto \sigma_p^3$.  Thus, the estimated richness $\lambda$ is a 
low-precision predictor of the measured values $\sigma_p$ or $M_{200}$
for individual clusters.

\begin{table}[th] \footnotesize
\begin{center}
\caption{\label{hecsredfits} \sc  Scaling Relations Between Dynamical Masses and Richnesses}
\begin{tabular}{lccc}
\tableline
\tableline
\tablewidth{0pt}
Relation & $b$   & $a$ & $\sigma_y$ \\
\tableline
$P(\sigma_p|\lambda)$ & $0.38^{+0.19}_{-0.18}$ & $0.050^{+0.024}_{-0.023}$ & 0.094$^{+0.018}_{-0.014}$ \\
extended sample & $0.240^{+0.047}_{-0.046}$ & $0.046\pm{0.011}$ & 0.090$^{+0.008}_{-0.007}$ \\
\tableline
$P(M_{200}|\lambda)$ & $0.99^{+0.46}_{-0.44}$ & $0.052\pm0.059$ & $0.252^{+0.044}_{-0.035}$ \\
extended sample & $0.58^{+0.14}_{-0.14}$ & $0.005^{+0.031}_{-0.032}$ & $0.276^{+0.021}_{-0.018}$ \\
\tableline
\end{tabular}
\end{center}
\tablecomments{Fits are of the relation $P(y|x)$ assuming 
the linear form $\log{y} = a + b\log{x}$ with intrinsic scatter $\sigma_{\log{y}}$ in the relation at fixed values of $\log{x}$.
Velocity dispersions are divided by 700$\kms$ and logarithmic masses are subtracted by 14.5. 
}
\end{table}

\subsection{Outliers}

The scaling relations show that many clusters are outliers from the main relation.  
Outliers can be produced by large overestimates or underestimates of either richness 
or velocity dispersion or by the intrinsic scatter in the scaling relation.  We use the scaling 
relation of $P(\sigma_p|\lambda)$ for the HeCS-red-ext sample to identify possible outliers.  
Specifically, we look at clusters that lie more than $\sigma_{int}$ away from this relation. 
Notably, all four outliers from HeCS-red with small $\sigma_p$ given their richness 
(A98S, RMJ0751, RMJ0830, RMJ2201) show a secondary peak at slightly higher 
redshift ($z_{secondary}-z_{red} < 0.04$, see Figures \ref{hecsredzhist1}-\ref{hecsredzhist2}) .  This result suggests that the richnesses of 
these clusters may be overestimated due to nearby background structure
included within the photometric redshift window. 

We provide additional information on some of the HeCS-red clusters below.

\begin{itemize}
\item{\em A98S}  This system is the southern component of the double 
cluster A98N/A98S \citep{forman81}.  The two X-ray peaks are 
separated by 8$\farcm$9 or 0.8$\Mpc$ on the sky.  Chandra observations show some 
evidence that A98N/A98S are in the early stages of merging \citep{paternomahler14}, while a 
third cluster, termed A98SS, lies 1.0$\Mpc$ South of A98S in projection.  \citet{beers82b} used redshifts 
of 24 member galaxies in A98N/A98S to study the system as a two-body merger
and conclude that a bound-ingoing model is most probable. 
Using a larger redshift sample, \citet{paternomahler14} confirm that 
A98N/A98S can be modeled as a two-body bound-ingoing 
model, although an unbound-outgoing scenario is also consistent with the data.  
They further report that A98SS is not bound to A98S.   
Although A98N has a larger X-ray luminosity than A98S \citep{forman81,jf99}, 
the X-ray temperature of A98N is only marginally larger than 
A98S \citep{paternomahler14}, and A98S contains more galaxies
than A98N. There is a bright galaxy at the center of both A98N and A98S. 
There are 39 (22) spectroscopically
confirmed members within 4$\farcm$4 of the bright galaxy at the center 
of A98S (A98N). The galaxies in A98N are separated by $-435\pm 252\kms$ from 
the galaxies in A98S.  The velocity difference from our larger sample 
is somewhat smaller than found by \citet{paternomahler14}.  This 
difference shifts the possible unbound-outgoing scenario for A98N/A98S 
to a bound-outgoing scenario.
Because the redshifts of A98N and A98S overlap, we treat the merging 
clusters A98N/A98S as a single massive system. 

With our larger redshift sample, we confirm that A98SS is separated 
from A98S by $\sim$5000$~\kms$ (visible near the top of the phase space diagram 
of Figure 4) and is therefore not gravitationally bound.
Members of A98SS produce the secondary peak in the redshift histogram of 
A98S visible in Figure 2.

\item{\em A115} A115 shows two prominent X-ray peaks, termed 
A115N/A115S, separated 
by 5$\farcm$1 or 1$\Mpc$ on the sky \citep{forman81}.  Chandra observations 
show that the intracluster medium (ICM) in the cores of A115N and A115S is cooler than 
the ICM in the outer parts, consistent with a subsonic merger \citep{gutierrez05}.  
\citet{barrena07} used 
a redshift survey to probe the dynamics of the cluster merger.  They found that 
A115N and A115S are separated by about 2000$\kms$ and that the velocity dispersion 
of A115S is somewhat larger than that of A115N.  They also find that a few galaxies 
are located at lower velocity and centered around a galaxy they term BCM-D. 
These galaxies are located outside the caustics in Figure \ref{hecsredc1}. 
We treat A115N and A115S as a combined system.

\end{itemize}

From the extended sample (HeCS-red-ext), some notable clusters are:

\begin{itemize}

\item{\em A963} A963 was noted by \citet{rozo14d} to be an outlier in the 
{\em Planck}-SZ-S/N-richness scaling relation.  A963 has the largest 
X-ray temperature for clusters for its richness.  \citet{rozo14d} inspected the DR8 
photometry around this cluster and concluded that there was a 
systematic uncertainty in the photometric background surrounding this 
cluster due to a bright star. The large uncertainties in estimated colors of 
galaxies may scatter many of them outside of the red sequence.  
\citet{andreon14} estimated the richness of A963 and found a much 
larger value than the redMaPPer algorithm.  Similarly,
HeCS \citep{hecsultimate,hwang14} showed that this cluster 
contains many spectroscopically confirmed members.  We show below 
($\S$ 4.4) that the richness estimate from our spectroscopy places this 
cluster much closer to the locus of points for the other clusters. 

\item{\em A1068} This cluster has a large velocity dispersion for its richness.  
A1068 was also noted as an outlier in the mass-richness relation by \citet{andreon14}.

\item{\em A1682} This cluster has a large velocity dispersion for its richness.

\item{\em MS0906 and A750} This pair of nearby clusters was noted in 
HeCS and found to have a large weak lensing mass for its caustic 
mass, presumably because the lensing mass includes both clusters 
while the caustic masses are able to separate the clusters \citep{hecslens}.  
Perhaps unsurprisingly, these clusters appear to be conflated by the 
redMaPPer algorithm as well, making them unusually rich for their 
velocity dispersions.

Using our spectroscopy to estimate the richnesses of these clusters, 
both clusters lie much closer to the main locus of points in both scaling relations
($\S 4.4$). 
This cluster pair is a good example of the ``catastrophic outliers" expected in 
richness-based cluster catalogs: two clusters of roughly comparable mass and 
richness that are also closely separated in redshift. Without spectroscopic 
redshifts, this cluster pair would be counted as a single cluster of roughly 
twice the mass of the individual clusters, resulting in a biased estimation 
of the cluster mass function. 

\item{\em A1758N and A1758S} Another pair of nearby clusters is A1758N and 
A1758S, each of which is composed of two merging clusters \citep{david04,okabe08,ragozzine12}. 
The redMaPPer algorithm identifies A1758N and A1758S as separate systems, 
but it does not detect their components.  Because the velocity distributions of 
the two clusters comprising A1758N overlap significantly, the measured 
velocity dispersion is probably not much larger than the velocity dispersions
of the individual clusters.  Thus, it is not surprising that the redMaPPer richness of the
two clusters comprising A1758N is large compared to its velocity dispersion.  

\end{itemize}

\section{Discussion}

\subsection{Reliability of Cluster Identification and Photometric Redshifts in the redMaPPer DR8 Catalog}

Our MMT/Hectospec observations of redMaPPer-selected clusters show that, 
at the relatively high richnesses of our target clusters ($\lambda \geq 64$), 
cluster candidates in the redMaPPer catalog correspond to real overdensities 
in redshift space.  Further, the redshift of the primary overdensity at the spatial 
position of the cluster candidate agrees well with the estimated redshift in the 
redMaPPer catalog. 

\subsection{Spectroscopic Completeness}

\begin{figure} 
\epsscale{1.0}
\plotone{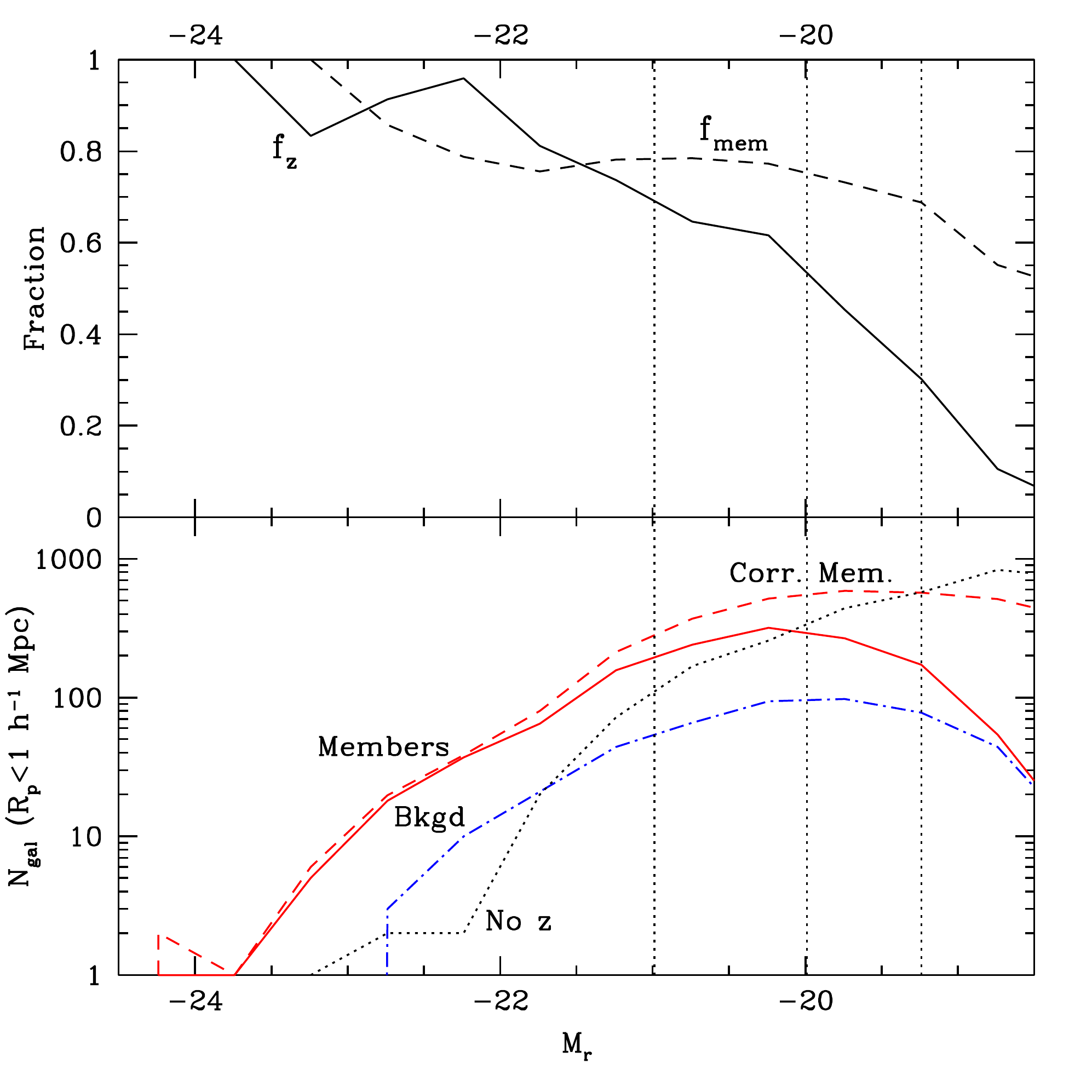}
\caption{\label{speccomp} (Top) Fraction of candidate red-sequence galaxies (projected radius within 1.0 Mpc and color within 0.1 mag of the red sequence) with spectroscopic redshifts as a function of fiducial absolute magnitude (i.e., non-members and unobserved galaxies are treated as if located at the distance of the cluster).
The dashed line shows the membership fraction of spectroscopically observed galaxies. 
Vertical dotted lines (from left to right) indicate absolute magnitudes of $M^*$, $M^*+1$, and $M^*+1.75$.
(Bottom) Number of candidate red-sequence galaxies  versus fiducial absolute magnitude.  Red solid, blue dash-dotted, and black dotted lines respectively show the number of spectroscopically confirmed members, spectroscopically confirmed background (and foreground) galaxies, and galaxies without redshifts.  The dashed line show the number of members after correcting for incompleteness. 
  }
\end{figure}

Figure \ref{speccomp} shows the spectroscopic completeness of the HeCS-red 
sample measured inside the Abell radius.  The upper panel shows the fraction of 
target galaxies (those within 0.1 mag of the red sequence) with redshifts and the
fraction of observed galaxies that are cluster members.  These fractions are displayed 
as functions of fiducial absolute magnitude (that is, the absolute magnitude that 
confirmed members or unobserved galaxies would have if they were cluster members). 
For conversion to absolute magnitudes, we use the fitting functions from 
\citet{westra10} based on empirical K-corrections from spectrophotometry 
with Hectospec and SDSS photometry.

The lower panel of Figure \ref{speccomp} shows the number of spectroscopically
confirmed members, confirmed background/foreground galaxies, and unobserved 
galaxies as a function of fiducial absolute magnitude. Vertical dotted lines 
indicate absolute magnitudes of $M^*$, $M^*+1$, and $M^*+1.75$.  
We categorize galaxies in these luminosity bins as bright, intermediate, and 
faint respectively.  The membership fraction of faint galaxies is similar 
to the membership fraction of bright galaxies.  The spectroscopic completeness 
decreases steadily with decreasing luminosity, but the completeness fraction 
of faint galaxies ($\sim$40\%) is sufficient to enable completeness 
corrections for the red-sequence luminosity function and cluster richness.
The dashed line in the lower panel of Figure \ref{speccomp} shows the 
total number of members in all HeCS-red clusters after correcting for incompleteness.  

\subsection{\label{memprobeval} Evaluation of Photometric Redshift Membership Probability }

Along with the cluster catalog, \citet{rykoff14} released a catalog of candidate cluster members 
where each galaxy has an assigned probability $P_{mem}$ that it is a member of a cluster.  
Here, we use our spectroscopic redshifts to assess the reliability of these probability estimates.

\begin{figure} 
\epsscale{1.0}
\plotone{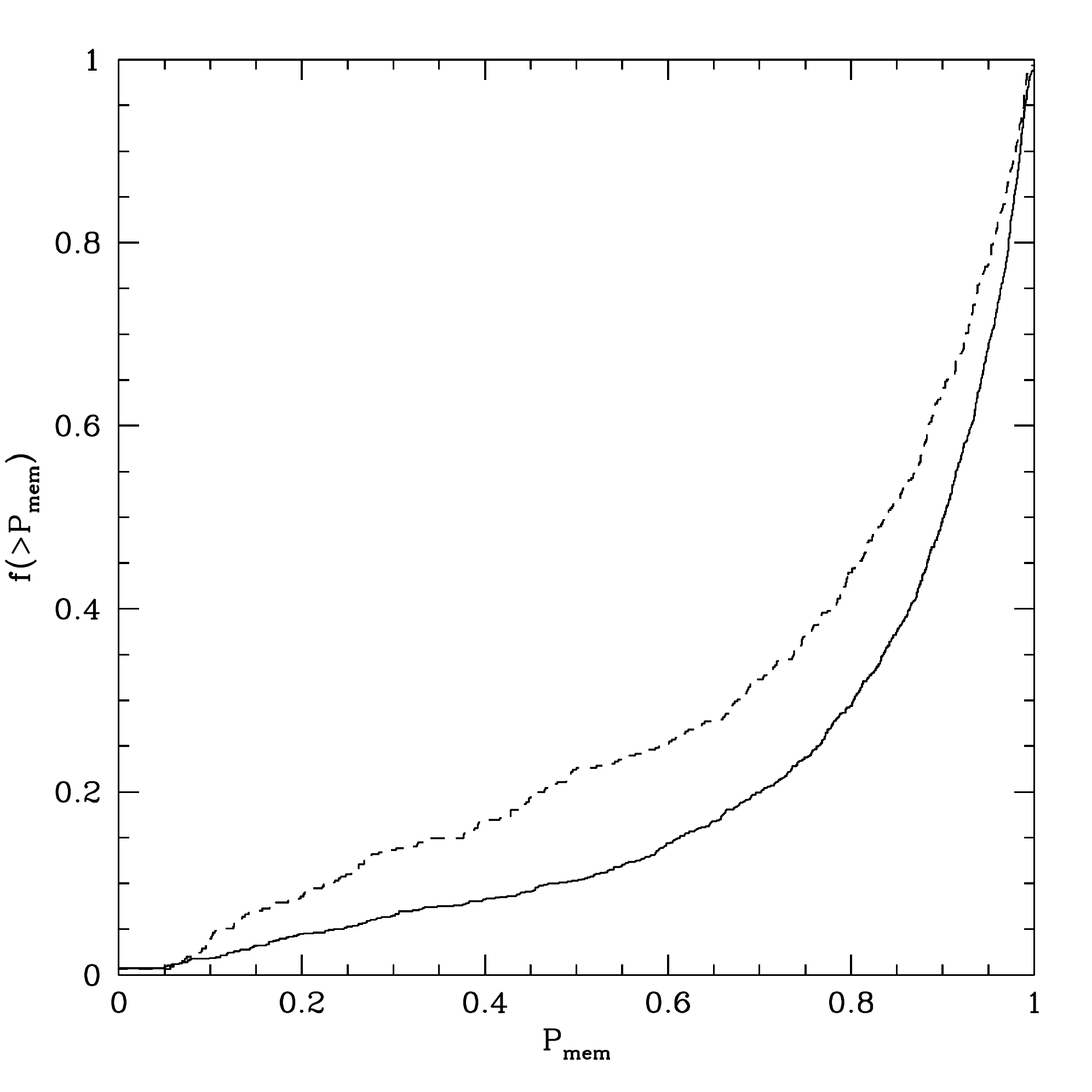}
\caption{\label{memprobks} Cumulative distributions of membership probability $P_{mem}$ 
for galaxies spectroscopically identified as clusters (solid line) or non-members (dashed line). 
  }
\end{figure}

We match our spectroscopic catalog to the redMapper membership catalog.  We identify 2159 
galaxies in common, of which 1710 are spectroscopically confirmed members (i.e., lie within 
the caustics).  Figure \ref{memprobks} shows cumulative distributions of $P_{mem}$ separated
by spectroscopic membership.  The median membership probability of confirmed members is 
$0.902\pm0.003$, and the median membership probability of confirmed non-members is 
$0.839\pm0.008$.  The high membership probability of confirmed non-members demonstrates 
that spectroscopic redshifts are required for identifying cluster members even for galaxies 
lying on the photometric red sequence. 

\begin{figure} 
\epsscale{1.0}
\plotone{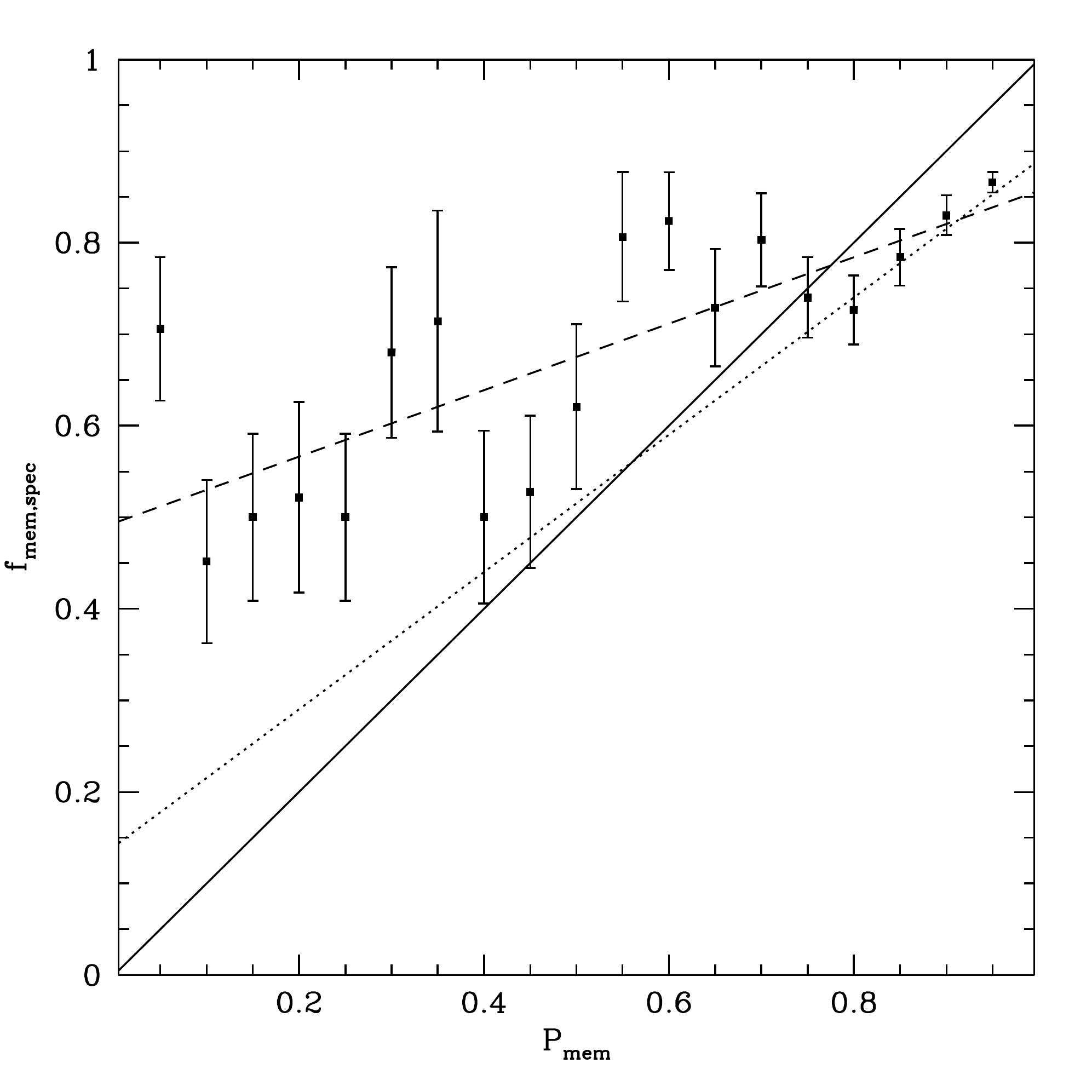}
\caption{\label{memprobspec} Spectroscopic member fractions of redMaPPer candidate members 
in bins of membership probability $P_{mem}$.  The solid line shows $f_{mem,spec}=P_{mem}$ 
and the dashed line shows an ordinary least squares fit to the data.  Uncertainties are 
computed assuming a binomial distribution.  The dotted line shows a similar comparison 
from analysis of (mostly smaller richness) redMaPPer clusters in the HectoMAP redshift survey (Sohn et al.~in prep.).
  }
\end{figure}

Figure \ref{memprobspec} quantifies this trend by calculating the spectroscopic membership 
fraction $f_{mem,spec} = N_{RMmem,spec}/N_{RMmem}$ as a function of $P_{mem}$. 
$N_{RMmem}$ is the number of redMaPPer candidate members with spectroscopic 
redshifts and $N_{RMmem,spec}$ is the number of redMaPPer 
candidate members classified as members with the caustic technique.  The member fraction 
is larger than $P_{mem}$ for all but the largest values of $P_{mem}$.   The dashed line in 
Figure \ref{memprobspec} shows an ordinary least squares fit to the data.  The equation for 
the line is $f_{mem,spec} = (0.361\pm0.097)P_{mem} + (0.496\pm0.056)$, indicating that 
$P_{mem}$ underestimates the actual membership probability at $P_{mem}<0.8$ and 
overestimates the membership probability at $P_{mem}>0.8$.

Our results are not consistent with \citet{rozo15b}, who conclude from a comparison of 
redMaPPer cluster candidates and GAMA redshifts that redMaPPer probabilities are largely
accurate with some small systematic effects. Their results are based on clusters with 
smaller richnesses ($\lambda <64$) than the HeCS-red clusters.  Sohn et al.~(in preparation)
perform a similar comparison using redshifts from the HectoMAP redshift survey
\citep{geller14,hwang16}.  The redMaPPer clusters in HectoMAP have smaller 
richnesses than HeCS-red clusters, and the membership fraction they measure 
(dotted line in Figure \ref{memprobspec}) is consistent with \citet{rozo15b}. 
Thus, the discrepant results on membership probabilities are consistent with 
a richness effect such that values of $P_{mem}$ of 0.05-0.7 from the redMaPPer algorithm 
underestimate the true membership probability for galaxies in high-richness clusters.

\begin{figure} 
\epsscale{1.0}
\plotone{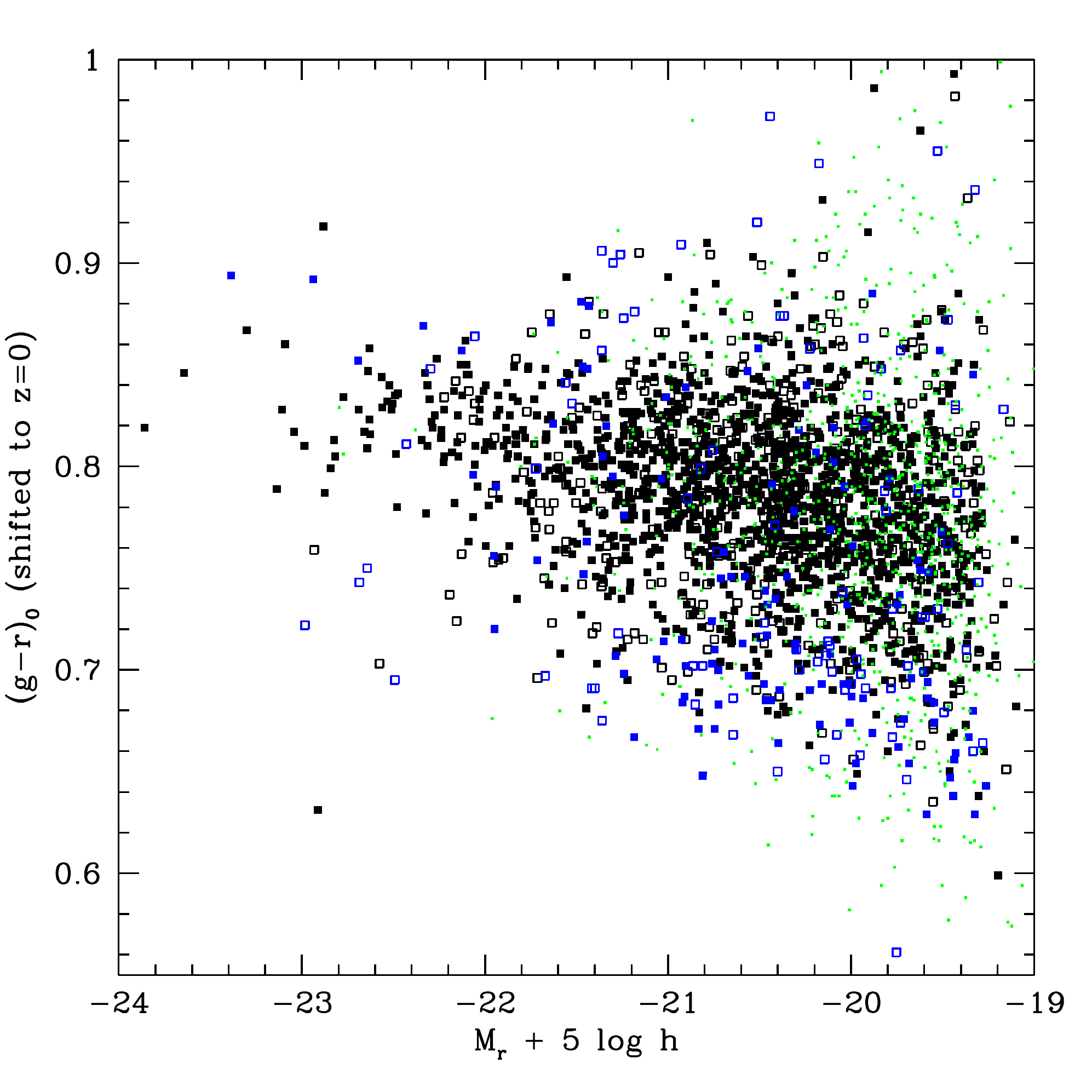}
\caption{\label{pmemrabsgr} Fiducial color-magnitude diagram for redMaPPer candidate members. Absolute magnitudes and colors are corrected to $z$=0 assuming that all galaxies lie at the redshift of the target cluster. Filled (open) points show spectroscopically identified members (non-members).  Blue points indicate galaxies with membership probability $P_{mem}<0.4$.  Small green points show redMaPPer candidate members without spectroscopic redshifts.    
  }
\end{figure}

Figure \ref{pmemrabsgr} shows a color-magnitude diagram for redMaPPer candidate members from 
the HeCS-red clusters.  For this figure, we plot fiducial colors and fiducial 
absolute magnitudes as if these galaxies were at the redshift of the cluster.   Candidates 
with $P_{mem}<0.4$ typically lie further away from the red sequence than 
candidates with larger $P_{mem}$.  Thus, the high spectroscopic membership fraction 
for galaxies with small $P_{mem}$ can largely be attributed to spectroscopic members that are 
somewhat faint and somewhat bluer than the red sequence.

\subsection{Richness Estimated Via Spectroscopy \label{specrichtext}}

Our spectroscopic data enable a test of the photometric richness parameter $\lambda$.
The richness parameter $\lambda$ is the estimated number of galaxies projected inside a 
cutoff radius that scales with the parameter $\lambda$.  That is, the cutoff radius is 
larger for clusters with larger $\lambda$. 

We adopt a simplified approach to estimate the richness of clusters using our spectroscopy. 
We use a fixed cutoff radius of $1 \Mpc$ for all clusters (this is the cutoff radius redMaPPer 
adopts for richness parameter $\lambda=100$).  We only include galaxies on the 
photometric red sequence.  We then measure the membership fraction in three bins of 
absolute magnitude M: brighter than $M^*$, $M^*<M<M^*+1$, and $M^*+1<M<M*+1.75$. 
Figure \ref{speccomp} shows that the HeCS-red spectra are $\sim$80\% complete in 
the brightest bin, $\sim$60\% complete in the intermediate bin, and $\sim$40\% 
complete in the faintest bin.  We correct for the incompleteness by dividing the number of 
members in each absolute magnitude bin by the completeness fraction of that bin.  We 
then sum the corrected counts to find the total number of cluster galaxies $N_{gal}$ within $1 \Mpc$ 
in projection.  We apply the same procedure to estimate the richnesses of two outliers in 
the HeCS-red-ext sample, A0963 and MS0906.  These clusters are no longer outliers 
in the $M_{200}-\lambda$ relation when we use our spectroscopic richness estimates 
(Figure \ref{m200vslambda}).

\begin{figure} 
\epsscale{1.0}
\plotone{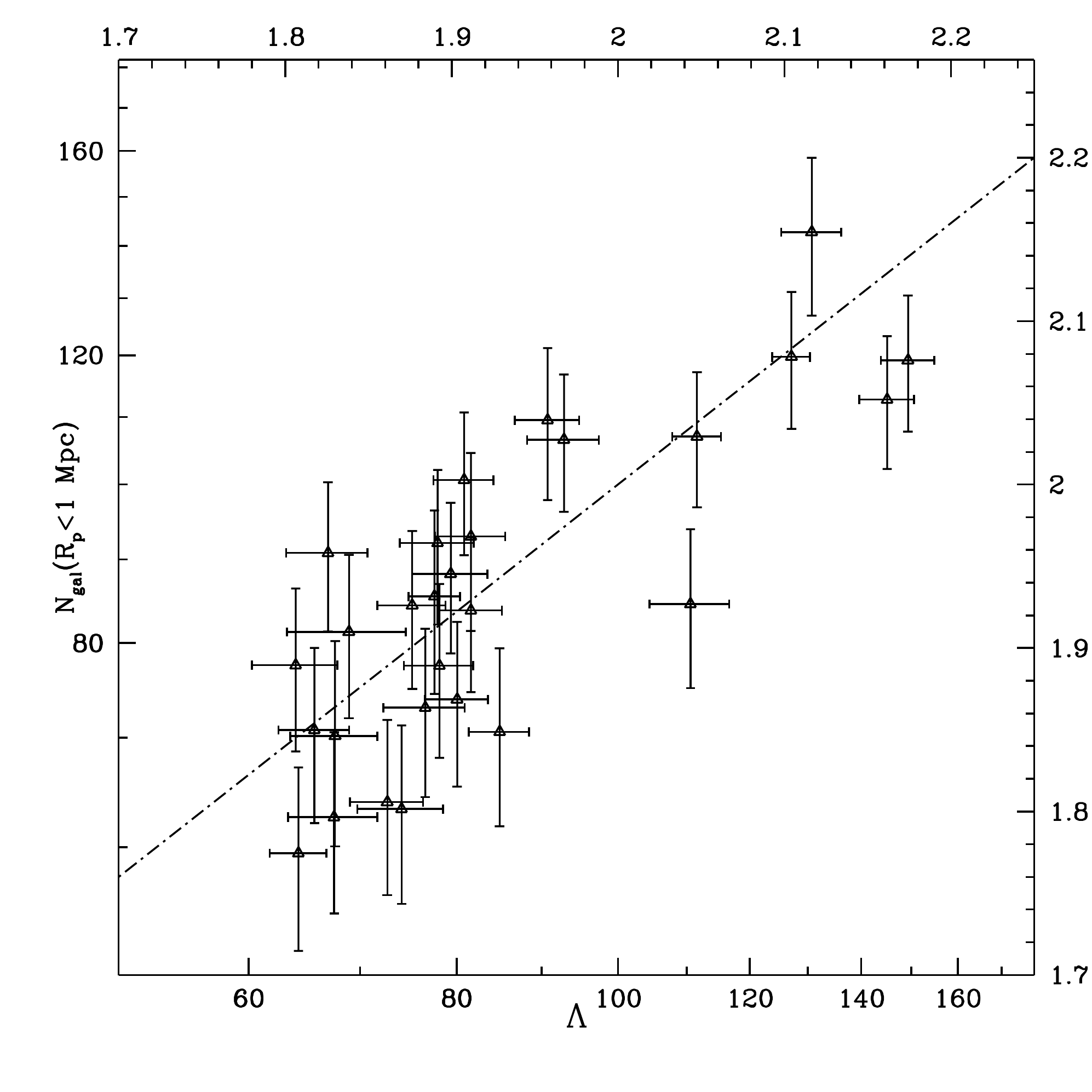}
\caption{\label{specrich} Richness parameter $\lambda$ versus
estimated red-sequence richness within $1\Mpc$ as estimated from 
spectroscopically classified members.  
The dash-dotted line shows a powerlaw with slope 0.8, corresponding to the slope 
expected if correcting the richness parameter $\lambda$ to a fixed aperture of $1\Mpc$ 
with a richness within $R_{ap}$ proportional to $R_{ap}$.  
 }
\end{figure}

Figure \ref{specrich} compares our spectroscopic richness estimates to the photometric 
redMaPPer richness estimates.  The two estimates of richness agree  with each other, 
suggesting that the photometric richness estimates from redMaPPer are closely related 
to each other. However, there is a large range in $N_{gal}$ (up to a factor of two) at
fixed $\lambda$.

One subtle aspect of the richness comparison is that the redMaPPer
algorithm does not use a fixed radial aperture to measure richness.  Instead, it 
uses a radial aperture that scales with the estimated richness; 
in practice, the aperture and richness parameters are fit jointly.  The redMaPPer radial 
aperture is given by $R_{ap,RM} = (\lambda/100)^{0.2}$.  For our redMaPPer-selected 
sample, $R_{ap,RM}$ changes by 18.5\% across the range of $\lambda$ covered by the 
sample.  We estimate the expected number of members within a fixed radial aperture of 1$\Mpc$
by assuming that the number of members increases approximately linearly 
with radius \citep[consistent with an NFW number density profile, see also Figure 16 of ][]{rykoff14}.
Under this assumption, the number of 
galaxies $\lambda_{1Mpc}$ expected within a fixed radial aperture of 1$\Mpc$ would be about 10\% larger
(smaller) than $\lambda$ for clusters with the smallest (largest) $\lambda$ in our sample. 
The dash-dotted line in Figure \ref{specrich} shows this prediction.  The agreement between 
the predicted relation and our measured relation shows that the average richness of 
redMaPPer clusters at fixed $\lambda$ is approximately equal to the richness $N_{gal}$ measured 
with spectroscopy.  However, the large range of $N_{gal}$ at fixed $\lambda$ indicates that 
there is significant uncertainty in this estimate for individual clusters.

\section{Conclusions}

We used spectroscopy from MMT/Hectospec to test the redMaPPer algorithm 
for detecting clusters and estimating richnesses. Our test is limited to 
high-richness, low-redshift systems where we can measure large numbers of redshifts
with MMT/Hectospec. 
We construct two samples, one selected purely on richness (HeCS-red), and 
a larger, sample that includes clusters selected on X-ray flux or SZ signal (HeCS-red-ext). 

We confirm that all of the cluster candidates 
in this high-richness sample are genuine clusters, although several 
systems show evidence of multiple structures along the line of sight.
The extended HeCS-red-ext sample shows evidence for bias in photometric redshifts 
of redMaPPer clusters, with $z_{spec}-z_{RM}=-0.0028\pm0.0005$.
This bias may be an effect of including background galaxies as cluster members when estimating photometric redshifts.

For the richness-selected HeCS-red sample, measured velocity dispersions 
correlate with cluster richness, and the 
scaling relation contains 24\% intrinsic scatter in velocity dispersion 
at fixed richness. Four clusters with small velocity dispersion given their 
estimated richness have background redshift peaks of nearby but unbound structure.
Nearby but unbound structure may produce overestimates of richness from photometric data.
The enlarged sample shows larger scatter and some outliers. 
Inspection of some of these outliers reveals that they can be 
caused both by problematic photometry (A963) and by `catastrophic' superpositions
of nearly equal-mass clusters (MS0906/A750). 
The range of velocity dispersion at fixed richness estimate $\lambda$ is a factor 
of two (three) for the HeCS-red (HeCS-red-ext) sample, and the range of 
measured $M_{200}$ at fixed $\lambda$ is a factor of roughly 10 (20). 
Thus, the richness estimate $\lambda$ is a low-precision predictor of 
$\sigma_p$ or $M_{200}$ for individual clusters. 

We compare spectroscopic membership classification to photometric membership
classification.  On average, the membership probability (estimated from photometry) of 
spectroscopically confirmed non-members is smaller than the 
membership probability of spectroscopically confirmed members. For the high-richness, 
low-redshift clusters in HeCS-red, the membership probabilities in redMaPPer appear 
to be underestimated for low-probability galaxies ($P_{mem}<0.75$)
and overestimated for high-probability galaxies ($P_{mem}\geq 0.8$). 

We estimate cluster richnesses from our spectroscopically determined member
catalogs.  The spectroscopic richnesses correlate well with the photometric 
richness estimates from redMaPPer.  Apparently, 
the underestimated probabilities for low-$P_{mem}$ candidate members 
are roughly balanced by the overestimated probabilities for high-$P_{mem}$
candidate members.
However, there is substantial scatter in the relation; the range of 
spectroscopic richness at fixed photometric richness is nearly a factor of two.  

Overall, our spectroscopic survey of red-sequence-selected clusters suggests 
that these cluster candidates are associated with significant overdensities in 
redshift space.  Photometric-based estimates of redshift, richness, and membership
probability correlate well with spectroscopic measures, although the scatter 
in individual objects is substantial.  Larger samples of red-sequence-selected 
clusters are necessary to provide robust constraints on the scaling relations 
between photometric richness and dynamical mass proxies like velocity dispersion.

HeCS-red focuses on low-redshift, high-richness clusters from the redMaPPer catalog.
RedMaPPer clusters cover the range 0.08 $< z <$ 0.6 and richnesses $\lambda > 20$. 
Thus, HeCS-red and HeCS-red-ext provide only a partial picture of the relationship 
between the spectroscopic properties of clusters and their redMaPPer counterparts. 
In a complementary paper, Sohn et al. test the full redshift and richness range of the 
redMaPPer catalog.

{\it Facilities:} \facility{MMT (Hectospec)}, \facility{FLWO: 1.5m (FAST)}

\acknowledgements

MJG is supported by the Smithsonian Institution.
AD acknowledges support from 
the INFN grant InDark.
JS acknowledges the support of a CfA Fellowship.
KR thanks the University of Washington for hosting a visit during which some of the work was performed.
We 
thank Susan Tokarz for reducing the spectroscopic data and Perry 
Berlind and Mike Calkins for assisting with the observations.  We also thank 
the telescope operators at the MMT and Nelson Caldwell for scheduling 
Hectospec queue observations. We thank the Telescope Data Center for assistance with data reduction.

Funding for
the Sloan Digital Sky Survey (SDSS) has been provided by the Alfred
P. Sloan Foundation, the Participating Institutions, the National
Aeronautics and Space Administration, the National Science Foundation,
the U.S. Department of Energy, the Japanese Monbukagakusho, and the
Max Planck Society. The SDSS Web site is http://www.sdss.org/.  The
SDSS is managed by the Astrophysical Research Consortium (ARC) for the
Participating Institutions. 
The Participating Institutions are The
University of Chicago, Fermilab, the Institute for Advanced Study, the
Japan Participation Group, The Johns Hopkins University, the Korean
Scientist Group, Los Alamos National Laboratory, the
Max-Planck-Institute for Astronomy (MPIA), the Max-Planck-Institute
for Astrophysics (MPA), New Mexico State University, University of
Pittsburgh, University of Portsmouth, Princeton University, the United
States Naval Observatory, and the University of Washington.


\bibliographystyle{apj}
\bibliography{rines}

\end{document}